\documentclass[prb,nobibnotes,altaffilletter,amsmath,amssymb,amsfonts]{revtex4}

\usepackage{inputenc}
\inputencoding{latin1}
\usepackage[T1]{fontenc}
\usepackage{graphicx}
\usepackage{palatino} 
\usepackage{mathpazo} 
\usepackage{bm} 
\usepackage{color}
\usepackage[usenames,dvipsnames]{xcolor}

\begin{document}

\title{Two measures of isochronal superposition}
\author{Lisa Anita Roed}
\author{Ditte Gundermann}
\author{Jeppe C. Dyre}
\author{Kristine Niss}
\address{DNRF Centre ``Glass and Time'', IMFUFA, Department of Sciences, Roskilde University, Postbox 260, DK-4000 Roskilde, Denmark } 
\date{\today}

\begin{abstract}
A liquid obeys isochronal superposition if its dynamics is invariant along the isochrones in the thermodynamic phase diagram (the curves of constant relaxation time). This paper introduces two quantitative measures of isochronal superposition. The measures are used to test the following six liquids for isochronal superposition: 1,2,6 hexanetriol, glycerol, polyphenyl ether, diethyl phthalate, tetramethyl tetraphenyl trisiloxane, and dibutyl phthalate. The latter four van der Waals liquids obey isochronal superposition to a higher degree than the two hydrogen-bonded liquids. This is a prediction of the isomorph theory, and it confirms findings by other groups.
\end{abstract}

\maketitle{}

The relaxation time of a supercooled liquid depends strongly on temperature and pressure. Different thermodynamic state points with same relaxation time are said to be on the same isochrone. A liquid obeys ``isochronal superposition'' if its dynamics is invariant along the isochrones. More precisely, a liquid obeys isochronal superposition (IS) if the complex, frequency-dependent response function $R(\omega,Q)$ in question at state point $Q$ can be written

\begin{equation}
R(\omega,Q)=R_{0}(Q)\tilde{R}(\omega,\tau(Q))+K_0(Q)\,.
\end{equation}
Here $R_{0}(Q)$ and $K_0(Q)$ are state-point dependent real constants. The function $\tilde{R}(\omega,\tau(Q))$ describes the shape of the relaxation spectrum; this function depends on the state point $Q$ only via its relaxation time $\tau(Q)$. For the imaginary part of the response function IS implies $R''(\omega,Q)=R_{0}(Q)\tilde{R}''(\omega,\tau(Q))$. This paper suggests two quantitative measures of how well IS is obeyed and applies them to test six glass-forming liquids for IS. 

T\"o{}lle (2001) first demonstrated IS for a single liquid, orthoterphenyl, in data for the intermediate scattering function determined by neutron scattering \cite{tolle01}. Soon after, systematic investigations of IS were initiated by Roland \textit{et al.} \cite{roland2003} and Ngai \textit{et al.} \cite{ngai2005} using dielectric spectroscopy. These seminal papers established IS for the van der Waals liquids studied, but reported that hydrogen-bonded liquids often violate IS. This was a striking discovery presenting a serious challenge to theory: Why would some liquids, for which the relaxation time spectrum generally varies throughout the thermodynamic phase diagram, have invariant spectra along its isochrones? And why do other liquids disobey IS?

By visually comparing the imaginary part of response functions along a liquid's isochrones the analysis of IS has traditionally followed the age-old method for investigating time-temperature superposition (TTS). The present paper takes the analysis one step further by suggesting two measures of the {\it degree of IS} --  such measures are relevant because one does not expect any liquid to obey IS with mathematical rigor. First, however, a few experimental details are given (further details are given in the supplementary material in the appendix).

\begin{table}
\begin{tabular}{|l|l|l|r|r|r|r|l|l|}
\hline
\tiny{Liquid} & \tiny{Abbr.} & \tiny{Bonding} & \tiny{$T_g$ (K)} & \tiny{$\Delta\varepsilon$} & \tiny{T (K)} & \tiny{p (MPa)} & \tiny{Density} & \tiny{Ref.} \\
\hline
\tiny{1,2,6 hexanetriol} & \tiny{1,2,6-HT} & \tiny{H} & \tiny{203 \cite{papini2012un}} & \tiny{$\sim40$}  & \tiny{236-251} & \tiny{100-400} & \tiny{This work} & \tiny{This work} \\
\tiny{Glycerol} &  & \tiny{H} & \tiny{193 \cite{nielsen2009}} & \tiny{$\sim60$} & \tiny{237-250} & \tiny{100-300} & \tiny{Ref. \onlinecite{bridgman1932}} & \tiny{This work} \\ 
\tiny{Polyphenyl ether} & \tiny{5PPE} & \tiny{vdW} & \tiny{245 \cite{jakobsen2005}} & \tiny{$\sim2$ \cite{jakobsen2005}} & \tiny{255-332} & \tiny{0.1-400} & \tiny{Ref. \onlinecite{gundermann2012}}  & \tiny{This work}\\
\tiny{Diethyl phthalate} & \tiny{DEP}  & \tiny{vdW} & \tiny{187 \cite{nielsen2009}} &  \tiny{$\sim8$} & \tiny{235-271} & \tiny{100-400} & \tiny{This work} & \tiny{This work}\\
\tiny{Tetramethyl tetraphenyl trisiloxane} & \tiny{DC704} & \tiny{vdW} & \tiny{211 \cite{jakobsen2005}} & \tiny{$\sim0.2$ \cite{jakobsen2005}} & \tiny{253-283} & \tiny{39-304} & \tiny{Ref. \onlinecite{gundermann2012}} & \tiny{Ref. \onlinecite{nielsen2010}} \\
\tiny{Dibutyl phthalate} & \tiny{DBP} & \tiny{vdW} & \tiny{177 \cite{niss2007}} & \tiny{$\sim8$} & \tiny{206-254} & \tiny{0-389} &  \tiny{Ref. \onlinecite{bridgman1932}} & \tiny{Ref. \onlinecite{niss2007}}\\
\hline
\end{tabular}
\caption{The liquids studied. "H" is hydrogen bonded and "vdW" is van der Waals bonded. Where no reference is given, $\Delta\varepsilon$ is from our measurements. Details on the density data are given in the supplementary material.}
\label{detailsliquids}
\end{table}

We studied four van der Waals liquids: polyphenyl ether (5PPE), diethyl phthalate (DEP), dibutyl phthalate (DBP), and tetramethyl tetraphenyl trisiloxane (DC704), as well as two hydrogen-bonded liquids: 1,2,6 hexanetriol (1,2,6-HT) and glycerol (details are given in table \ref{detailsliquids}). New measurements have been obtained for four of the liquids, while the data on DC704 are from Ref. \onlinecite{nielsen2010} and the data on DBP are from Ref. \onlinecite{niss2007}. The four liquids studied in this work were all studied before, also by use of dielectric spectroscopy under high pressure \cite{forsman1988,forsman1986,paluch1996,gundermann2012,paluch1997,pawlus2003}. 

Glycerol was dried in an exicator for 20 hours and 1,2,6-HT was dried for two hours; the other liquids were used as acquired from Sigma Aldrich. The experiments were performed on the high-pressure equipment described in Refs. \onlinecite{gundermann2012} and \onlinecite{roed2012}. The electrical measurement equipment is described in Ref. \onlinecite{igarashi2008}. 
Pressures go up to 600 MPa, temperatures range from 233 to 333 K. The sample cell consists of two round stainless steel plates with a diameter of 19.5 mm separated by a 0.05 mm thick Kapton spacer, which has inner diameter 17.5 mm and outer diameter 19.5 mm. To document reproducibility all measurements were repeated. The relaxation time was identified from the dielectric loss-peak frequency.

\begin{figure}
\begin{minipage}{0.3\textwidth}
\centering
\includegraphics[width=1\textwidth]{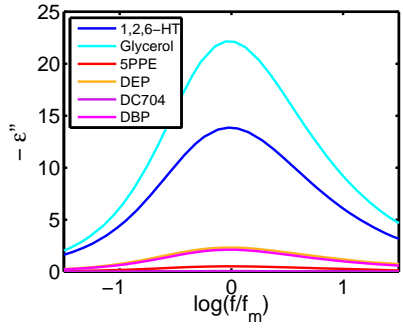}
\end{minipage}
\caption{One dielectric relaxation spectrum for each of the six liquids plotted on a linear scale. The hydrogen-bonded liquids (blue) have much larger signals than the van der Waals liquids (reddish). The spectrum of DC704 is too small to be visible.}
\label{figure1}
\end{figure}

Hydrogen-bonded liquids have generally much larger dipole moment than van der Waals liquids, which leads to much better dielectric relaxation signals for the former liquids. Figure \ref{figure1} shows typical dielectric relaxation data for the six liquids studied.

To develop quantitative measures of IS consider first perfect IS. A loss peak is characterized by, in principle, infinitely many shape parameters $X,Y,Z,...$. In practice a few parameters are enough to characterize the shape, for instance fitting-model based parameters like the stretching exponent $\beta$, the Havriliak-Negami parameters, the Cole-Davidson $\beta_{CD}$, etc, or model-independent parameters like the half width at half depth or the loss peak area in a log-log plot. Perfect IS is characterized by constant shape parameters along an isochrone: $0= dX|_\tau = dY|_\tau = dZ|_\tau=...$, in which $|_\tau$ signals that the variation is considered at constant relaxation time. 

In order to determine how much a given shape parameter $X$ varies along an isochrone, we assume that a metric $ds$ has been defined in the thermodynamic phase diagram. Since $d\ln X =dX/X$ gives the relative change of $X$, the rate of relative change of $X$ along an isochrone is given by the operator $L$ defined by

\begin{equation}\label{Ldef}
L(X)\equiv \left|\frac{d\ln X}{ds} \right|_\tau\,.
\end{equation}
How to define a reasonable metric $ds$? A state point is characterized by its temperature $T$ and pressure $p$, so one option is $ds^2=dT^2+dp^2$. This metric depends on the unit system used, however, which is not acceptable. This problem may be solved by using logarithmic distances, i.e., defining $ds^2=(d\ln T)^2+(d\ln p)^2$. Actually, using pressure is not optimal because there are numerous decades of pressures below ambient pressure where little change of the physics take place; furthermore, a logarithmic pressure metric does not allow for negative pressures. For these reasons we instead quantify state points by their temperature and density $\rho$, and use the following metric

\begin{equation}\label{dsdef}
ds^2\equiv (d\ln T)^2+(d\ln \rho)^2\,.
\end{equation}
Equations (\ref{Ldef}) and (\ref{dsdef}) define a quantitative measure of IS. In practice, suppose an experiment results in data for the shape parameter $X$ along an isochrone. This gives a series of numbers $X_1,X_2,...$, corresponding to the state points $(T_1,\rho_1), (T_2,\rho_2),... $. Since $\ln X_{i+1} - \ln X_i =\ln(X_{i+1}/X_i)$, etc, the discrete version of the right-hand side of Eq. (\ref{Ldef}) is 
$\left|\ln (X_{i+1}/ X_i)/{\sqrt{\ln^2(T_{i+1}/T_i)+\ln^2(\rho_{i+1}/\rho_i)}} \right|$.

An alternative measure of IS corresponding to the purely temperature-based metric $ds^2=(d\ln T)^2$ is defined by 

\begin{equation}\label{LTdef}
L_T(X)\equiv \left|\frac{d\ln X}{d \ln T} \right|_\tau\,.
\end{equation}
This measure is considered because density data are not always available. If the density-scaling exponent $\gamma\equiv (d\ln T/d\ln \rho)|_\tau$ is known for the range of state points in question, Eq. (\ref{dsdef}) implies $ds^2=(d\ln T)^2(1+1/\gamma^2)$. This leads to the following relation between the two IS measures 

\begin{equation}\label{relation}
L(X)=\frac{L_T(X)}{\sqrt{1+1/\gamma^2}}\,.
\end{equation}
This relation is useful in the (common) situation where gamma is reported in the literature but the original density data are difficult to retrieve.

As one shape parameter we used the half width at half depth, $W_{1/2}$, defined as the number of decades of frequency from the frequency of  maximum dielectric loss to the higher frequency (denoted $f_{1/2}$) where the loss value is halved: $W_{1/2}\equiv\log(f_{1/2}/f_{m})$ \cite{nielsen2009}. Here and henceforth ``$\log$`` is the logarithm with base 10. 

\begin{figure}
\begin{minipage}{0.3\textwidth}
\centering
\includegraphics[width=1\textwidth]{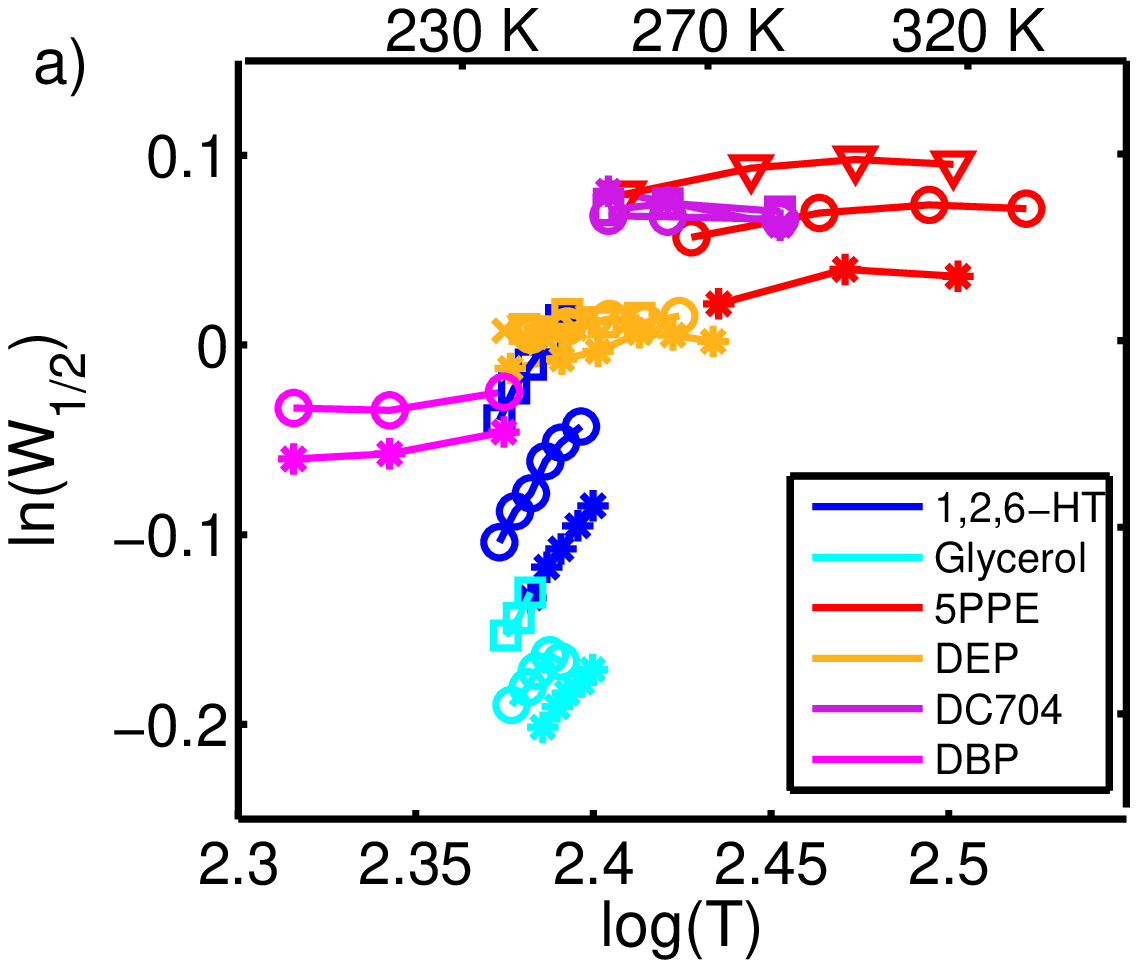}
\end{minipage}
\begin{minipage}{0.3\textwidth}
\centering
\includegraphics[width=1\textwidth]{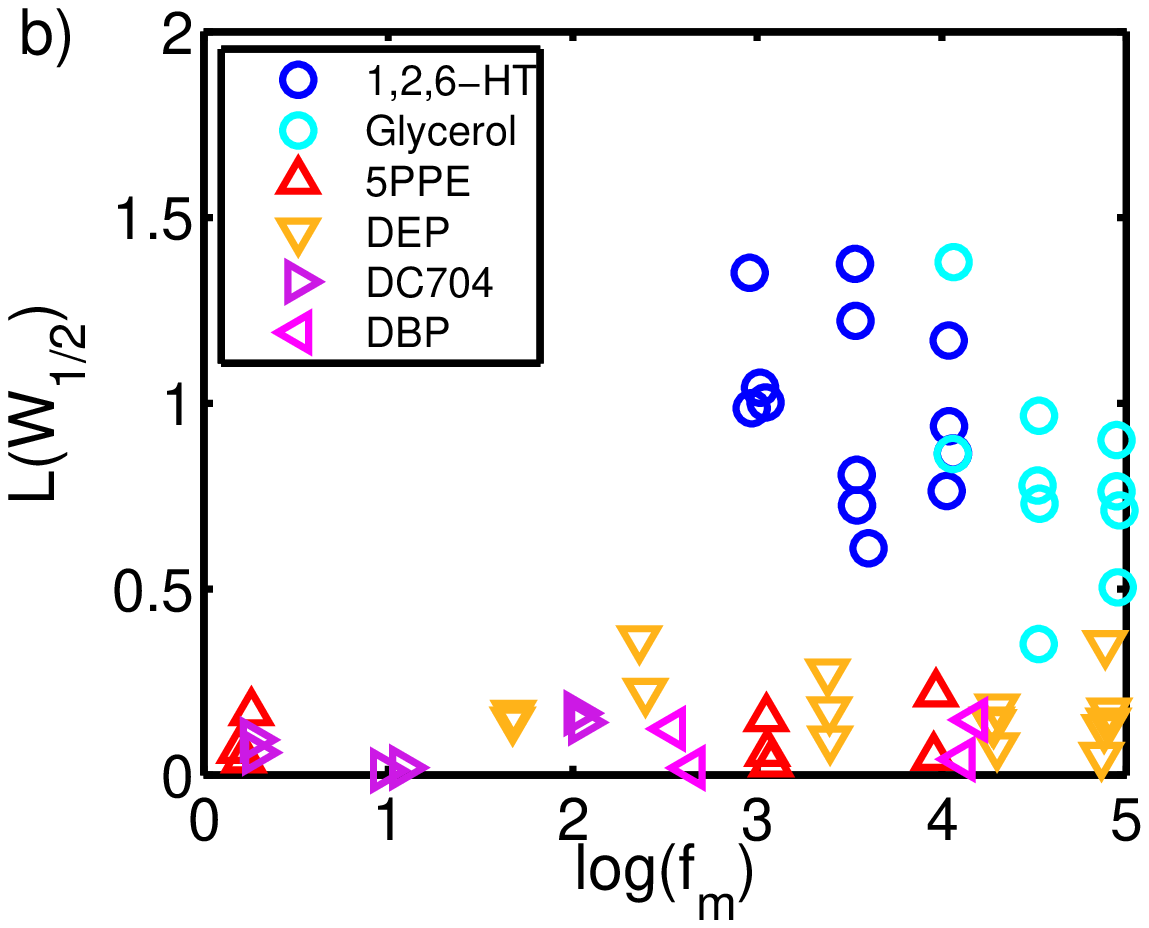}
\end{minipage}
\begin{minipage}{0.3\textwidth}
\centering
\includegraphics[width=1\textwidth]{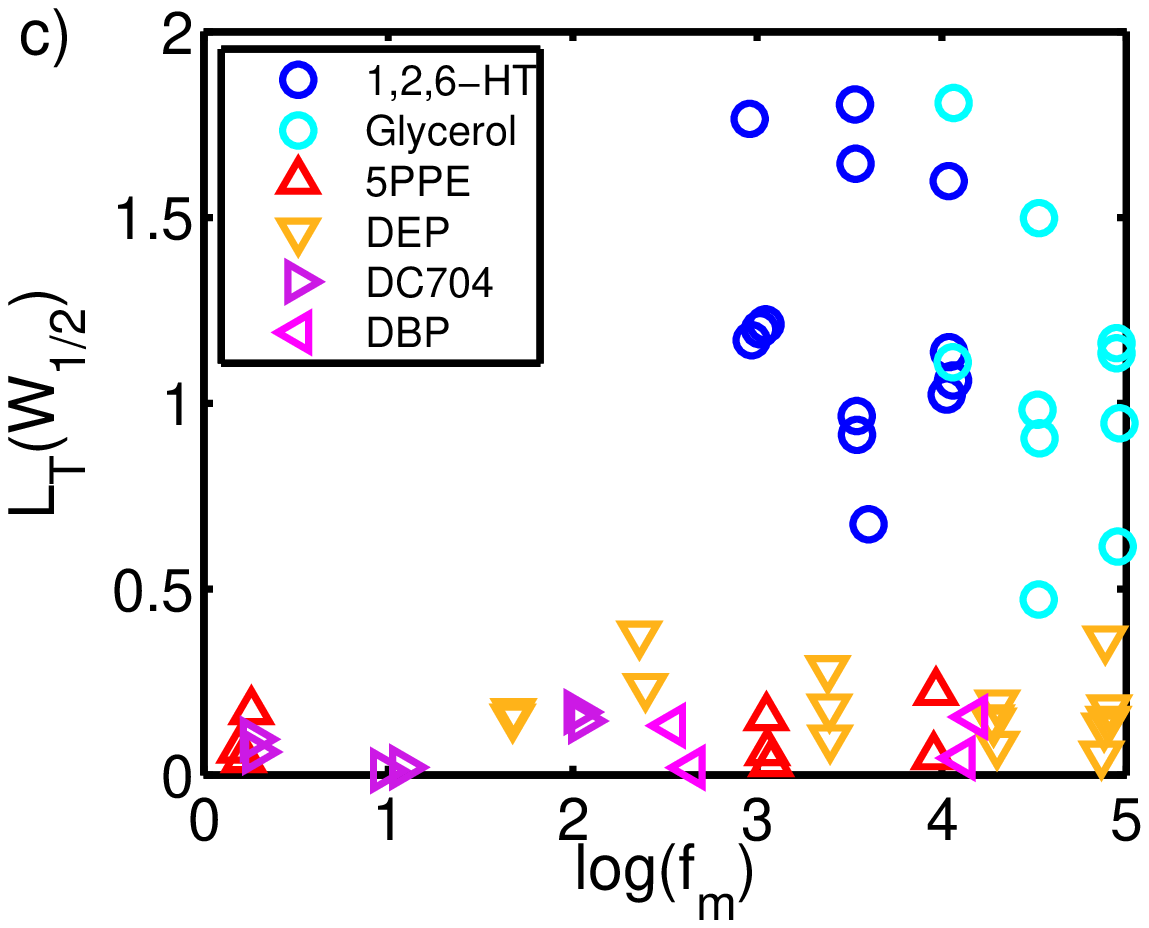}
\end{minipage}
\caption{Analysis of isochronal superposition (IS) based on the half width at half depth of the dielectric loss peak, $W_{1/2}$.
(a) Data for $W_{1/2}$ along isochrones marked by connecting lines as functions of temperature.
(b) The measure $L(W_{1/2})$ giving the rate of relative change of $W_{1/2}$ along an isochrone. 
(c) The measure $L_T(W_{1/2})$. 
Both measures are considerably higher for the hydrogen-bonded liquids than for the van der Waals liquids; thus the latter obey IS to a higher degree than the hydrogen-bonded liquids.}
\label{figure4}
\end{figure}

The $W_{1/2}$-values are plotted in Fig. \ref{figure4}(a) as functions of temperature for each isochrone studied. This figure suggests that IS is not obeyed for the two hydrogen-bonded liquids (blue), but for the four van der Waals liquids. There is more scatter in the values for the van der Waals liquids than for the hydrogen-bonded liquids, which is due to the smaller dielectric signals (see Fig. \ref{figure1}).

From Fig. \ref{figure4}(a) it is seen that the $W_{1/2}$ is almost constant at
all state points for DEP and DC704, suggesting that IS in these cases
is a consequence of TTS (or more specifically,
time-temperature-pressure superposition). However, the isochrones in
Fig. \ref{figure4}(a) are separated from one another in the case of 5PPE and DBP,
just as they are in the case of 1,2,6-HT and glycerol. This shows that
IS can apply even when the spectra broaden upon
supercooling, which is also supported by earlier works where IS has been found for systems without TTS \cite{nielsen2010,capaccioli2007}.

For a quantitative IS analysis we apply the $L$ and $L_T$ operators to the shape parameter $W_{1/2}$ (Figs. \ref{figure4}(b) and (c)). From these figures it is clear that the measures for the hydrogen-bonded liquids are higher than for the van der Waals liquids. It is not straight forward to estimate the systematic uncertainties involved in the measurements, but an attempt to do so is presented in the supplementary material. We see from Fig. \ref{figure4} that the two measures $L(W_{1/2})$ and $L_T(W_{1/2})$ lead to similar overall pictures and the same conclusion.

\begin{figure}
\begin{minipage}{0.3\textwidth}
\centering
\includegraphics[width=1\textwidth]{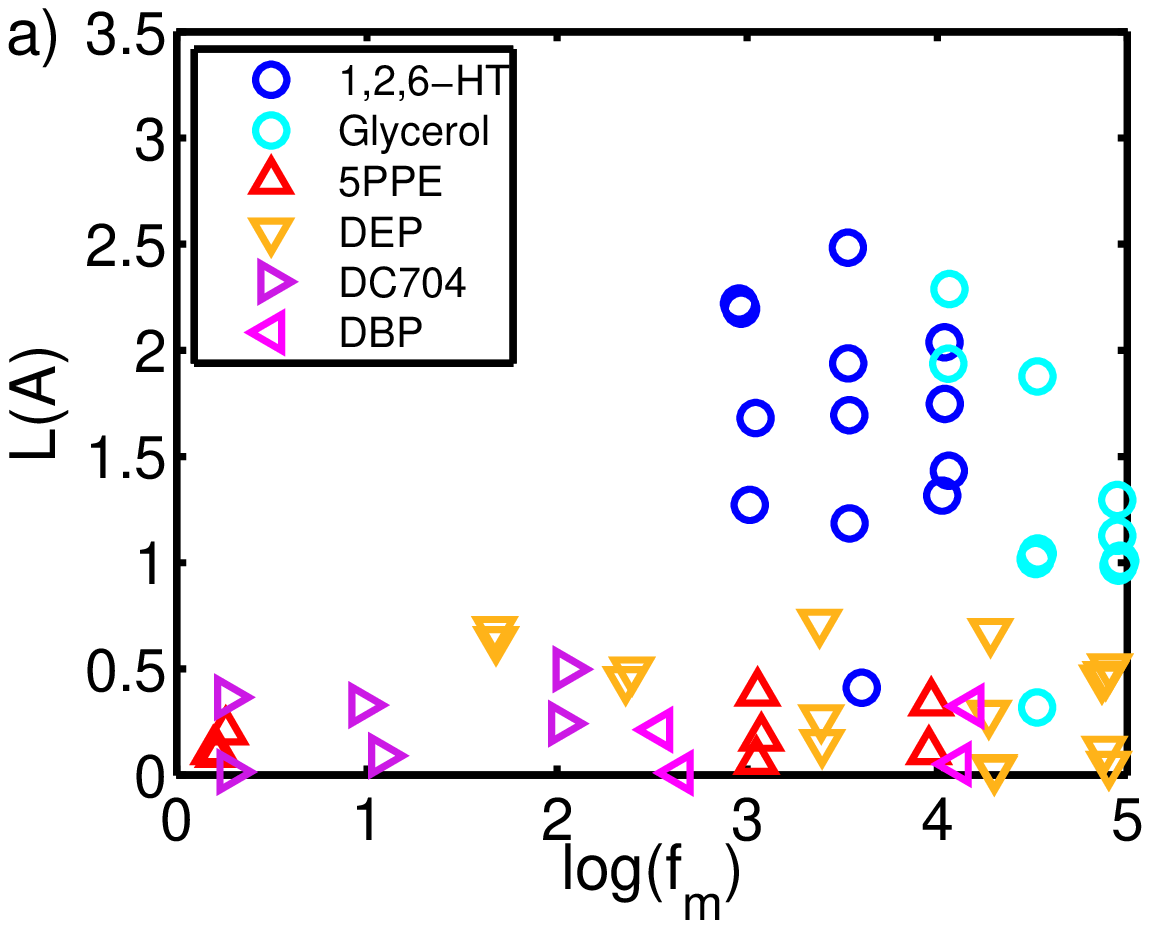}
\end{minipage}
\begin{minipage}{0.3\textwidth}
\centering
\includegraphics[width=1\textwidth]{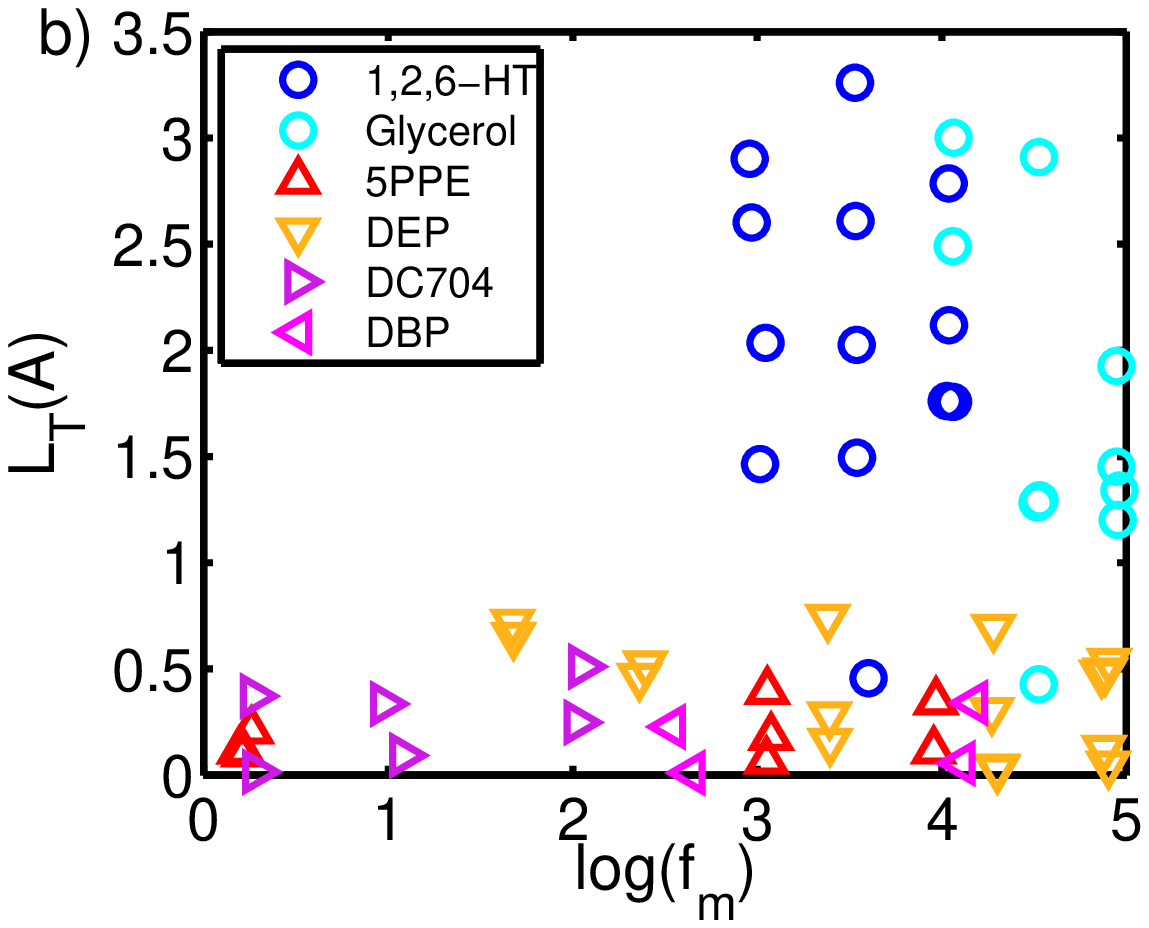}
\end{minipage}
\caption{The measures of isochronal superposition based on the area $A$ of the dielectric loss peak plotted in a normalized log-log plot, obtained by integrating from -0.4 to +1.0 decades around the loss-peak frequency. 
(a) The measure $L(A)$.
(b) The measure $L_T(A)$.}
\label{figure5}
\end{figure}

Invariance of one shape parameter like $W_{1/2}$ is not enough to prove IS. As a second, model-independent shape parameter we used the area of dielectric loss over maximum loss in a log-log plot, denoted by $A$. We integrated from -0.4 decades below the loss peak frequency to 1.0 decade above it. This was done by adding data for the logarithm of the dielectric loss taken at -0.4, -0.2, 0.2, 0.4, 0.6, 0.8, and 1.0 decades relative to the loss peak frequency. This area measure focuses on the high-frequency side of the peak. This is motivated in part by the occasional presence of dc conductivity on the low-frequency side of the peak, in part by the fact that for molecular-liquids this side of the peak is generally characterized by a slope close to unity, i.e., it varies little from liquid to liquid  \cite{nielsen2009phd}. In Fig. \ref{figure5} the measures $L(A)$ and $L_T(A)$ are shown. Clearly, the measures are higher for the hydrogen-bonded liquids than for the van der Waals liquids. Thus also with 
respect to 
the 
area shape parameter, the van der Waals liquids obey IS to a higher degree than the hydrogen-bonded liquids.

\begin{figure}
\begin{minipage}{0.3\textwidth}
\centering
\includegraphics[width=1\textwidth]{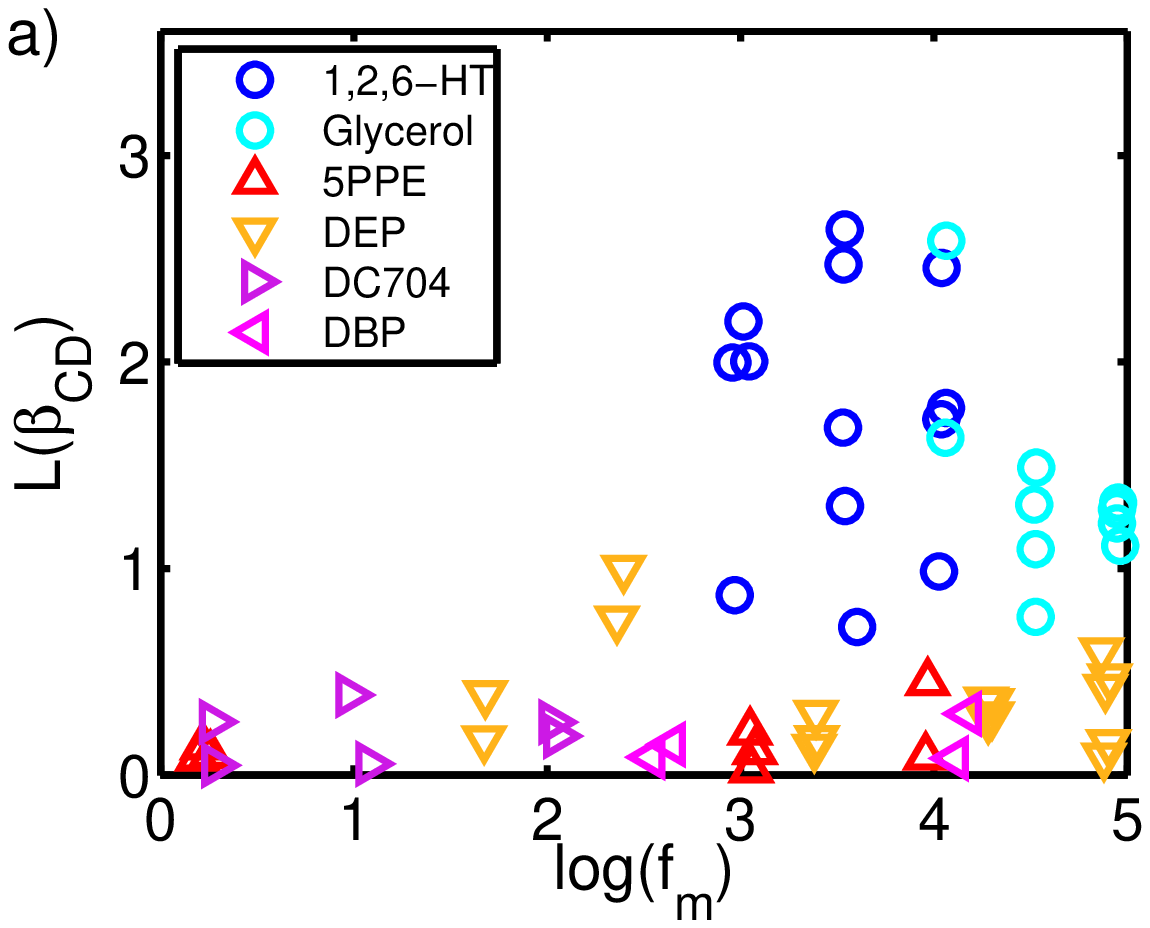}
\end{minipage}
\begin{minipage}{0.3\textwidth}
\centering
\includegraphics[width=1\textwidth]{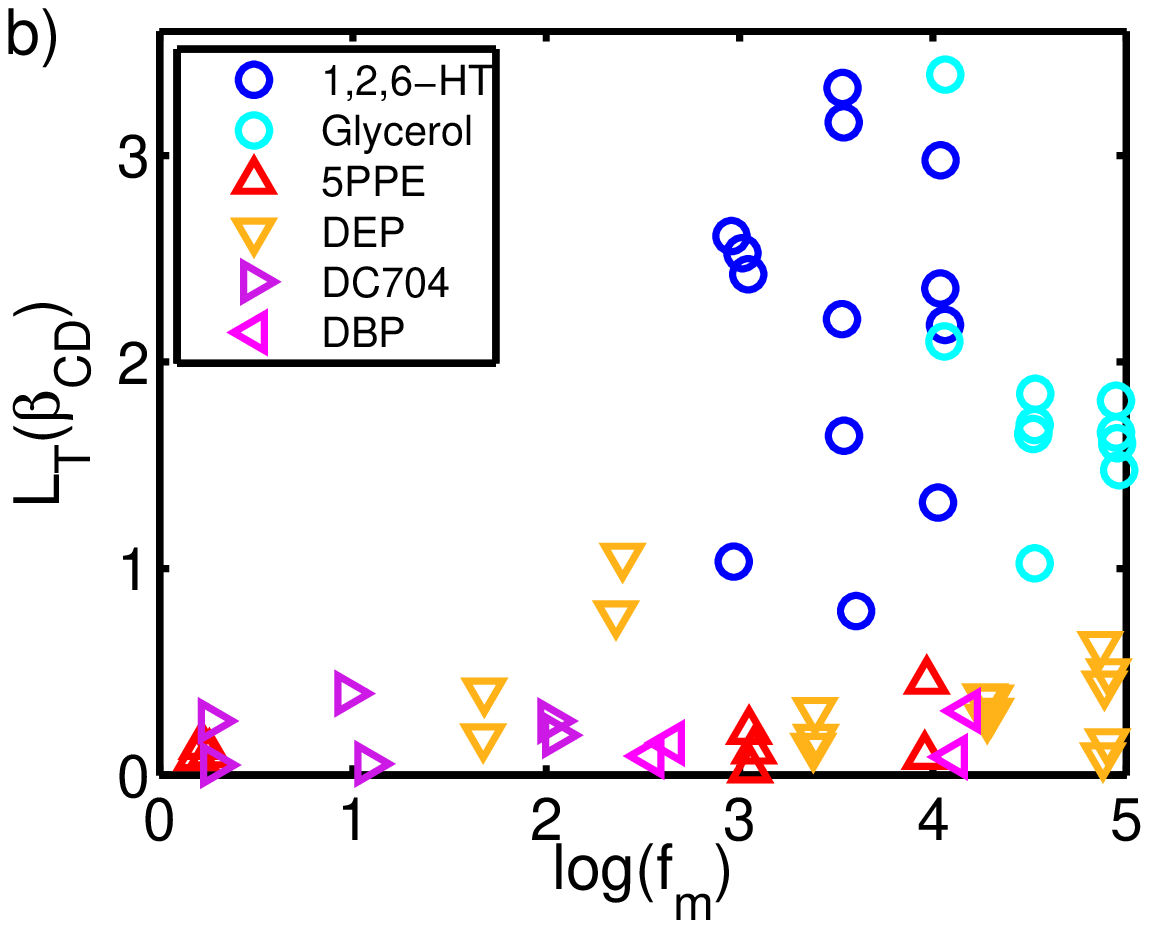}
\end{minipage}
\caption{The measures of isochronal superposition based on the shape parameter of the Cole-Davidson function $\beta_{CD}$. 
(a) The measure $L(\beta_{CD})$.
(b) The measure $L_T(\beta_{CD})$.}
\label{figure6}
\end{figure}

As a third parameter, we used the model-dependent shape parameter $\beta_{DC}$ from the Cole-Davidson fitting function \cite{cole1950}. Details on the fits are given in the supplementary material. Results from using the L operators on $\beta_{CD}$ are shown in Fig. \ref{figure6}. It is seen that also with respect to the model-dependent shape parameter $\beta_{CD}$ the van der Waals liquids obey IS to a higher degree than the hydrogen-bonded liquids. 

The conclusion that van der Waals liquids obey IS better than hydrogen-bonded liquids is not new. This was reported as a clear tendency in the pioneering papers on IS from 2003 and 2005 by Roland \textit{et al.} \cite{roland2003} and Ngai \textit{et al.} \cite{ngai2005}. In 2009 this finding was given a theoretical basis via the isomorph theory \cite{gnan2009,schrder2011}, which applies for liquids that have strong correlations between their virial and potential-energy equilibrium fluctuations at constant volume\cite{bailey2008a,bailey2008b,schrder2009}. Due to the directional nature of hydrogen bonds, liquids dominated by these do not show strong virial potential-energy correlations \cite{bailey2008a}. The isomorph theory predicts that density scaling, as well as IS, applies for van der Waals liquids, but not for hydrogen-bonded liquids, a result that is consistent with previous experiments  \cite{dreyfus2004,casaliniroland2004,alba2004,roland2008}. Only few experimental 
studies have yet been made with the explicit purpose of testing the isomorph theory, see e.g. Ref. \onlinecite{gundermann2011}, which supports the theory. 

In this paper we have focused on systems with no visible
beta relaxation or excess wing. Capaccioli \emph{et al.}  \cite{capaccioli2007}
demonstrated IS in systems with beta relaxation, suggesting that the beta and alpha relaxations are
connected. This result can be rationalized in terms of the isomorph theory,
which predicts IS to hold for all intermolecular modes. 
It would be interesting to use the
quantitative measures suggested above for
systems with a beta relaxation.

To summarize, we have proposed two measures of isochronal superposition that quantify the relative change of a given relaxation-spectrum shape parameter along an isochrone. The measures were tested for six liquids with regard to two model-independent shape parameters characterizing dielectric loss peaks, the half width at half depth and the loss-peak area in a normalized log-log plot, and one model-dependent shape parameter, the Cole-Davidson $\beta_{CD}$. The two measures lead to similar overall pictures; in particular both support the conclusion that van der Waals liquids obey IS to a higher degree than hydrogen-bonded liquids. 

We thank Albena Nielsen for providing data to this study, Bo Jakobsen for help with the data analysis, and Ib H\o{}st Pedersen for help in relation to the high-pressure setup. L. A. R. and K. N. wish to acknowledge The Danish Council for Independent Research for supporting this work. The center for viscous liquid dynamics ``Glass and Time'' is sponsored by the Danish National Research Foundation's Grant No. DNRF61. 

\section*{Appendix: Supplementary}

\tableofcontents
\subsection{Challenges in the investigation}

\begin{figure}
\begin{minipage}{0.35\textwidth}
\centering
\includegraphics[width=1\textwidth]{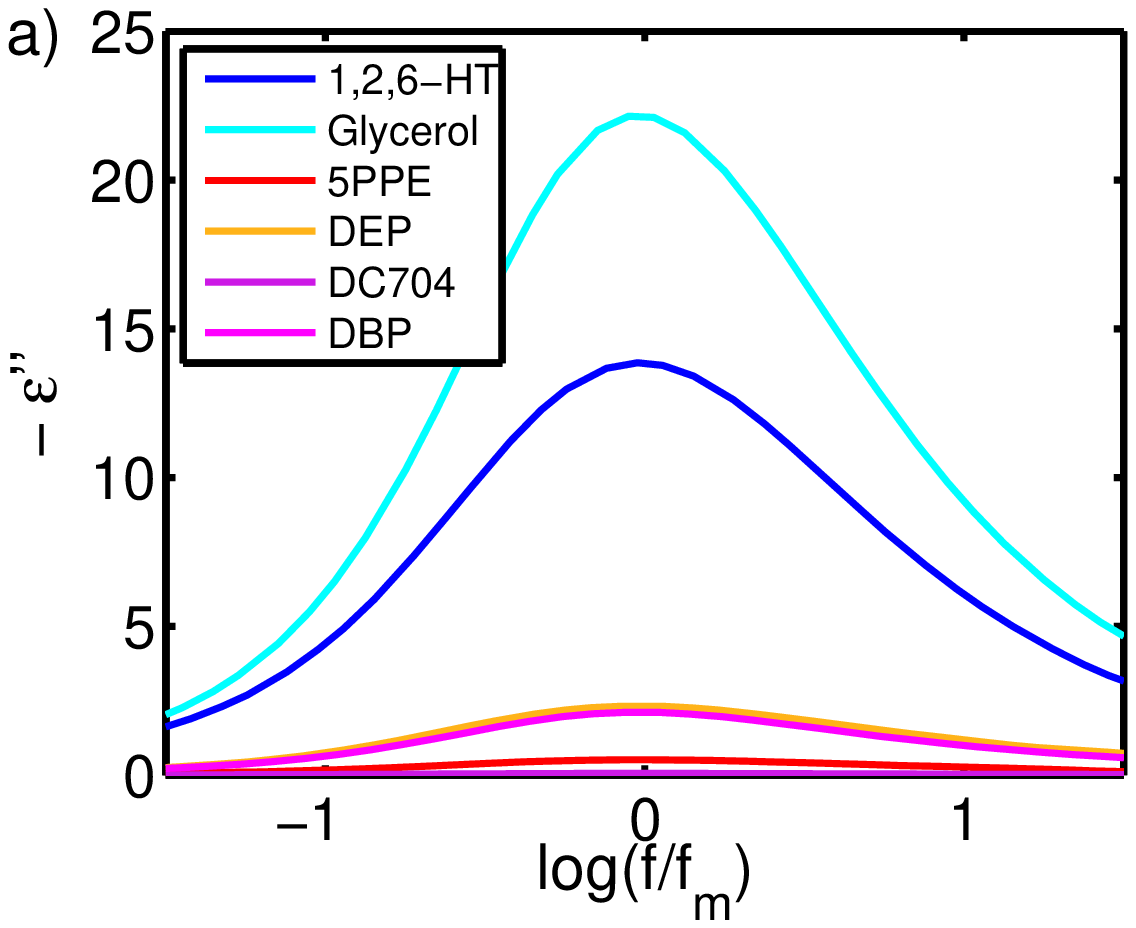}
\end{minipage}
\begin{minipage}{0.35\textwidth}
\centering
\includegraphics[width=1\textwidth]{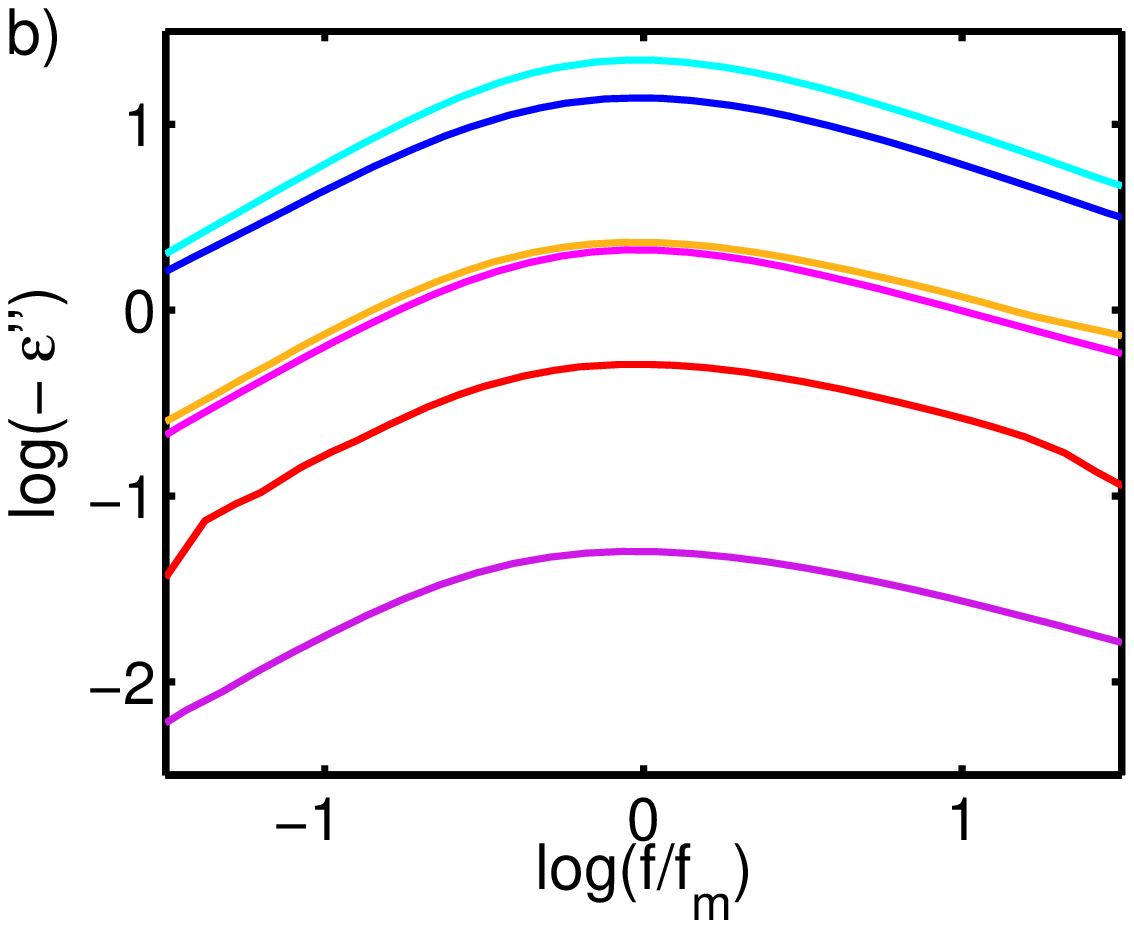}
\end{minipage}
\begin{minipage}{0.35\textwidth}
\centering
\includegraphics[width=1\textwidth]{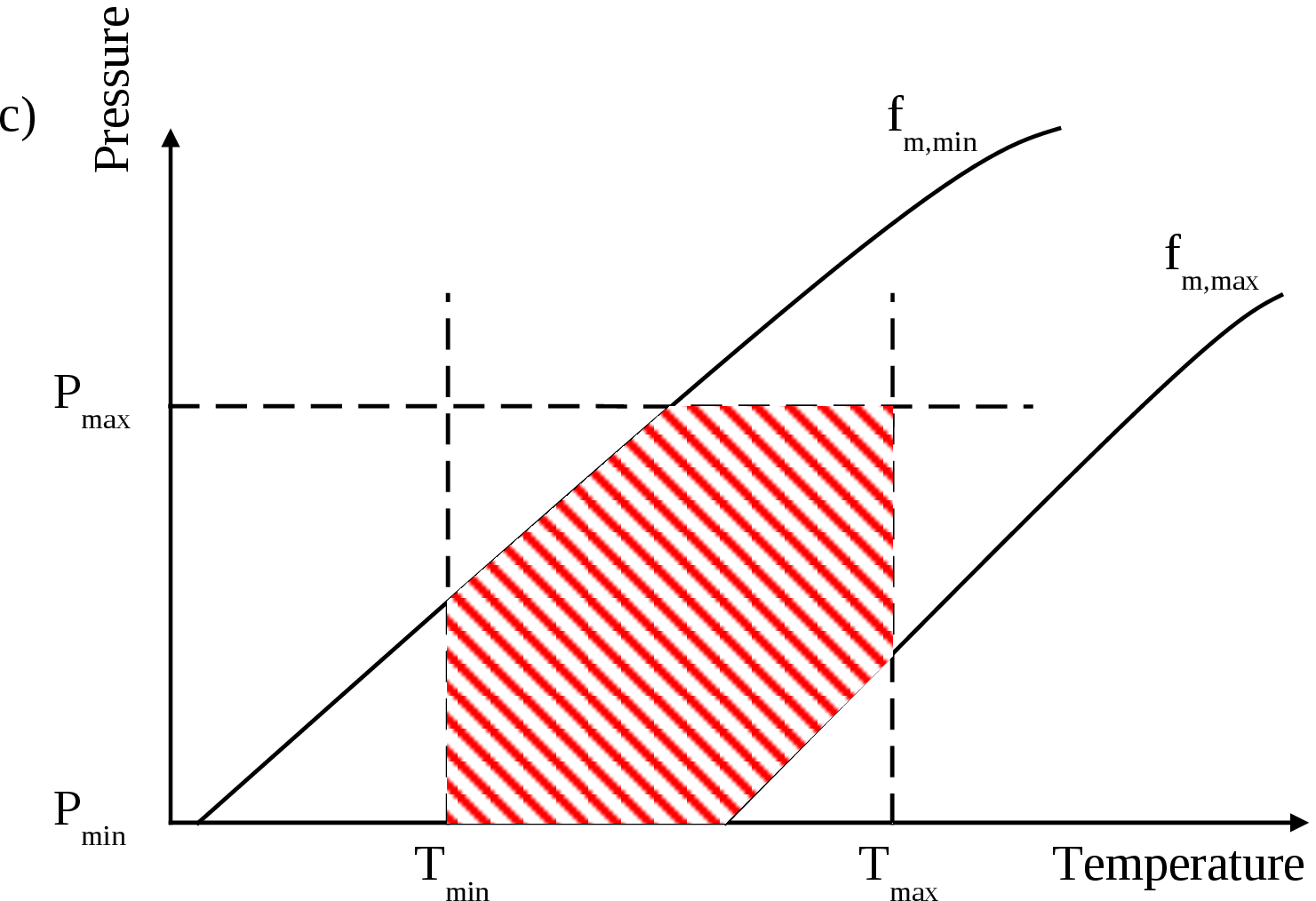}
\end{minipage}
\begin{minipage}{0.35\textwidth}
\centering
\includegraphics[width=1\textwidth]{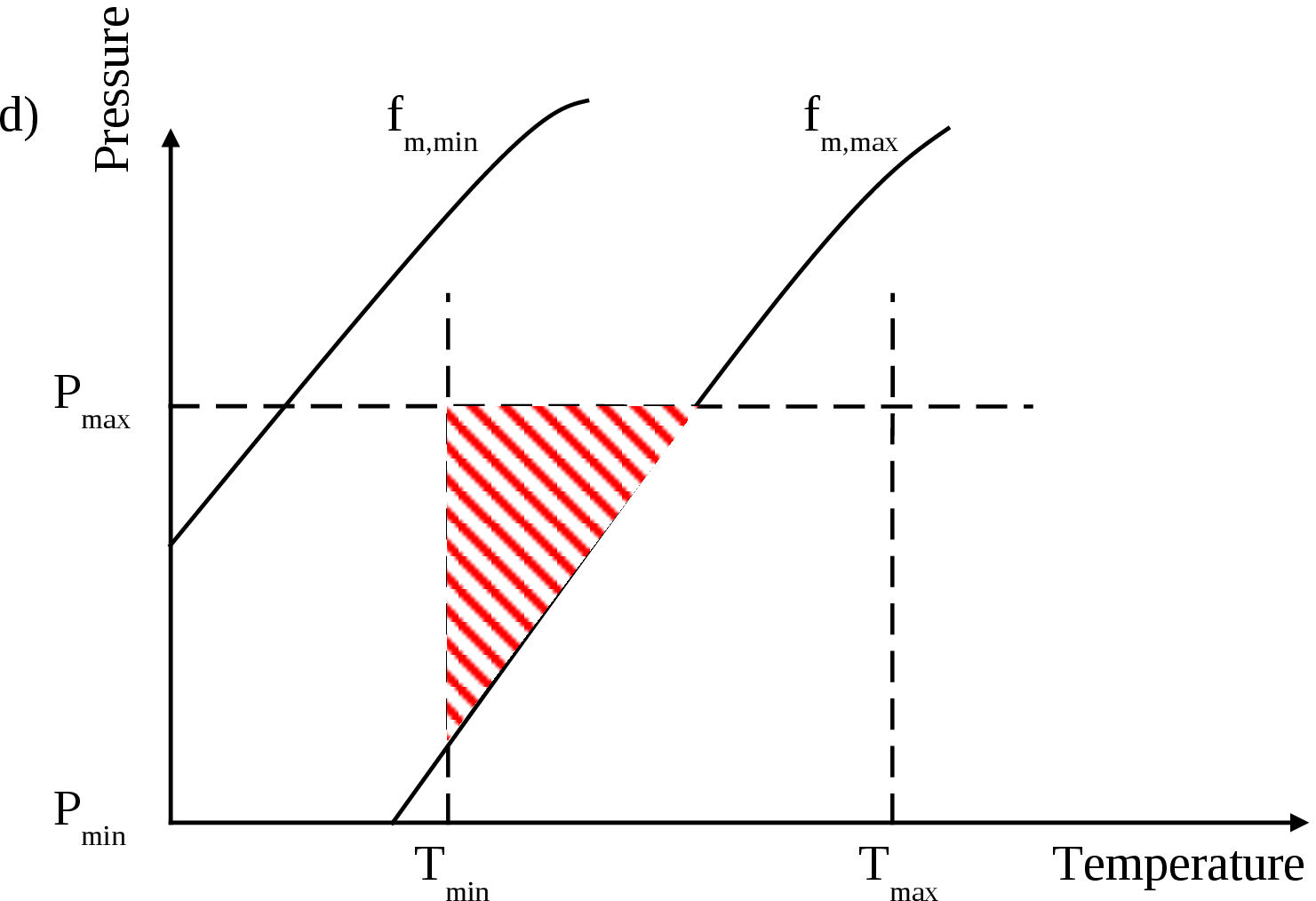}
\end{minipage}
\caption{These figures illustrate the challenges in relation to an
  investigation of isochronal superposition. The figures (a) and (b)
  show one relaxation spectrum for each liquid on respectively a
  linear and a log scale. It is seen, that the hydrogen bonded liquids
  (the blue spectra) have better signal than the van der Waals
  liquids. The spectrum of DC704 is not visible in figure (a), because the
  signal is too low compared to the other liquids. In the figures (c)
  and (d), the red shaded area illustrates the part of the phase
  diagram where the $\alpha$-peak can be measured.  The pressure,
  temperature, and frequency range are determined by the
  equipment. However, the frequency at the $\alpha$-relaxation
  ($f_{m}$) at each state point depends on the liquid. Two
  liquids are illustrated: the glass transition temperature $T_g$ is
  lower for the liquid in (d) than the liquid in (c).}
\label{figure1s}
\end{figure}

In relation to an investigation of isochronal superposition there is several things to consider. 
When using dielectric spectroscopy the amplitude of the signal is strongly sample dependent. 
Figure \ref{figure1s} illustrates the size of the signal for
the liquids studied in this work and the overall experimental situation.

Another factor which differs from sample to sample is the area in the
phase diagram where measurements can be performed. The equipment has a
temperature, a pressure, and a frequency range, which gives some
general limits. The temperature and pressure dependence of the
relaxation time is sample dependent, and the part of the phase diagram
where the $\alpha$-relaxation is present in the available frequency
range therefore varies. Figure \ref{figure1s} illustrates this. 

\subsection{Experimental details}

The pressure vessel (of the type MV1) and the pump are from Unipress
Equipment in Warsaw, Poland. The temperature is controlled by a thermal bath with a temperature
stability of $\pm0.02$ K in the bath \cite{julabo2009}. A thermocouple
measures the temperature close to the sample cell, and this is the
temperature used in the data treatment. The sample cell is wrapped in Teflon tape
and rubber to avoid mixing with the pressure fluid.

Further information about the high-pressure equipment can be seen in Refs. \onlinecite{gundermann2012} and \onlinecite{roed2012} and further information about the electrical measurement equipment can be seen in Ref. \onlinecite{igarashi2008}. 

Measurements on DEP and 5PPE were performed at several temperatures
along respectively 7 and 5 isobars. With the liquids 1,2,6-HT and
glycerol, we discovered crystallization, which was detected by a drop
in the dielectric signal. For these liquids, we therefore 
performed the final measurements with a new sample directly on selected isochrones. 

\subsection{Raw data}
\begin{figure}
\begin{minipage}{0.34\textwidth}
\centering
\includegraphics[width=1\textwidth]{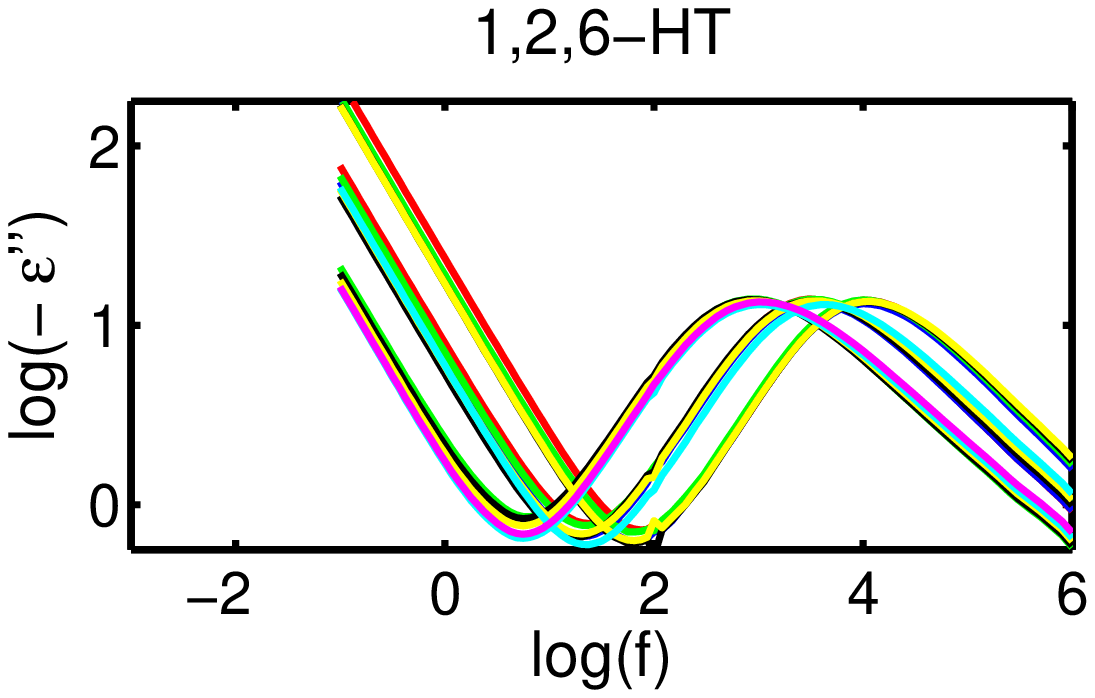}
\end{minipage}
\begin{minipage}{0.34\textwidth}
\centering
\includegraphics[width=1\textwidth]{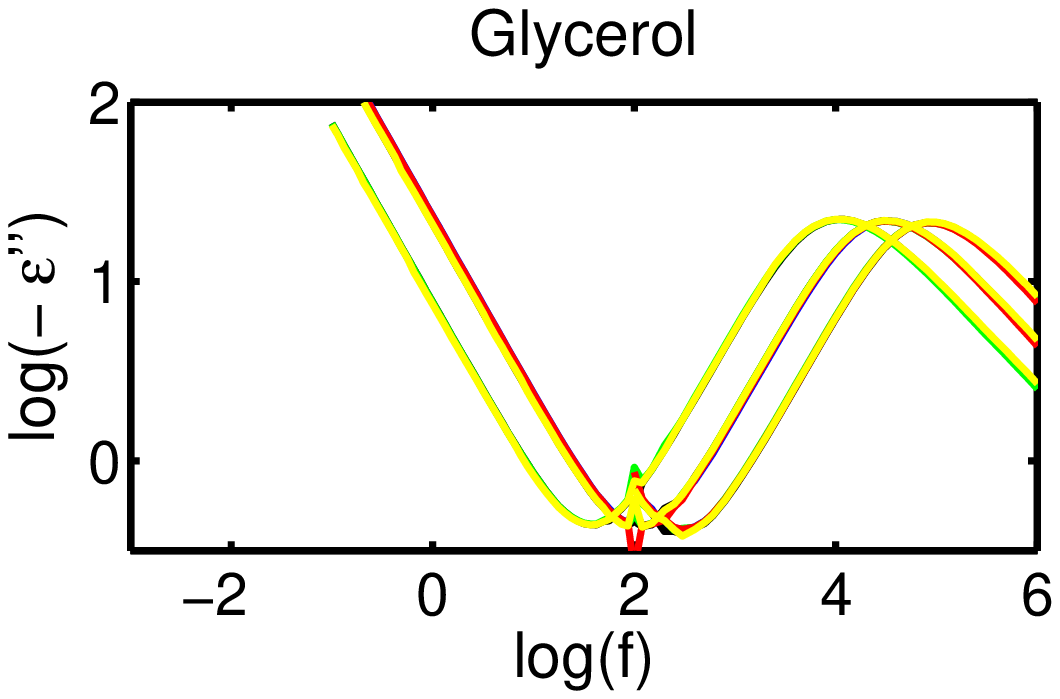}
\end{minipage}
\begin{minipage}{0.34\textwidth}
\centering
\includegraphics[width=1\textwidth]{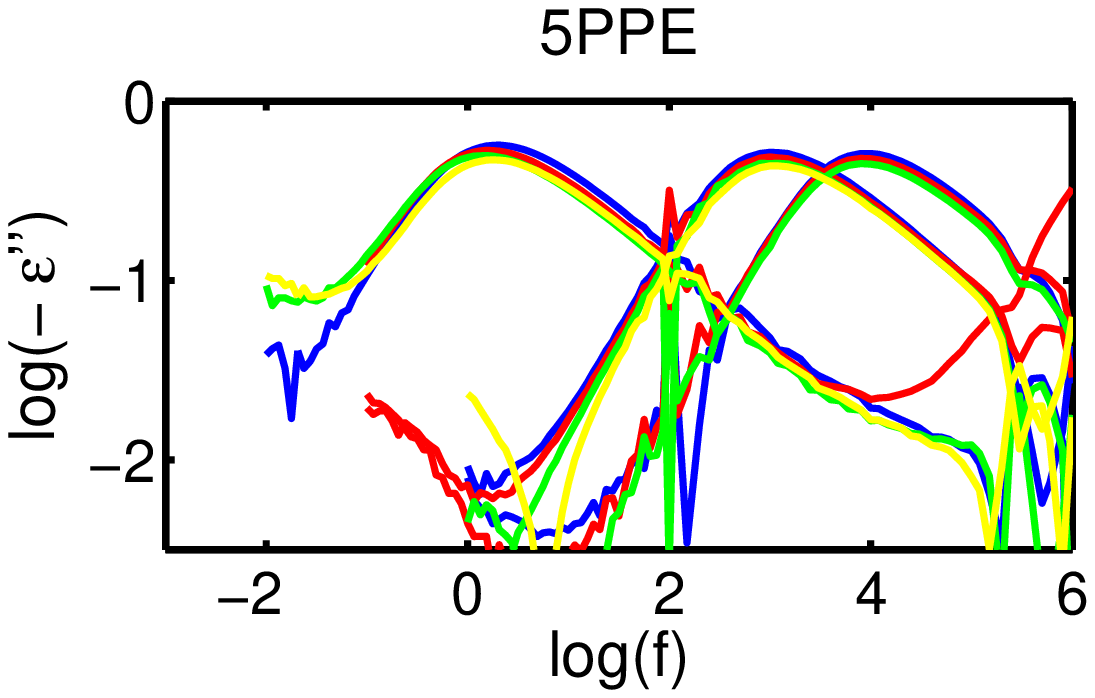}
\end{minipage}
\begin{minipage}{0.34\textwidth}
\centering
\includegraphics[width=1\textwidth]{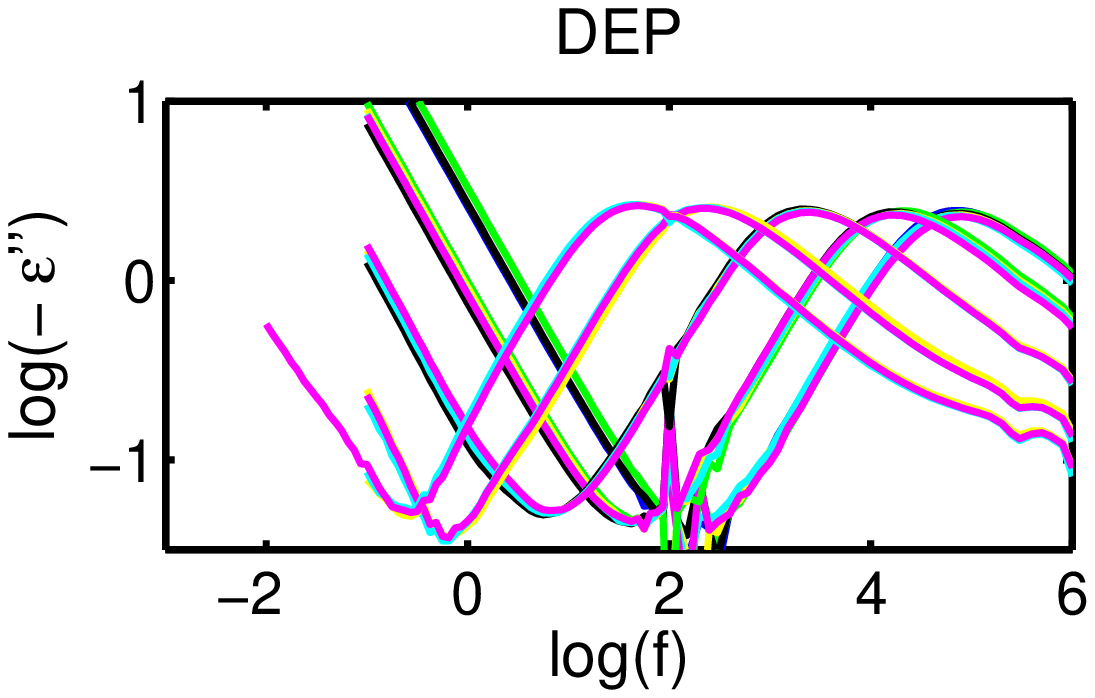}
\end{minipage}
\begin{minipage}{0.34\textwidth}
\centering
\includegraphics[width=1\textwidth]{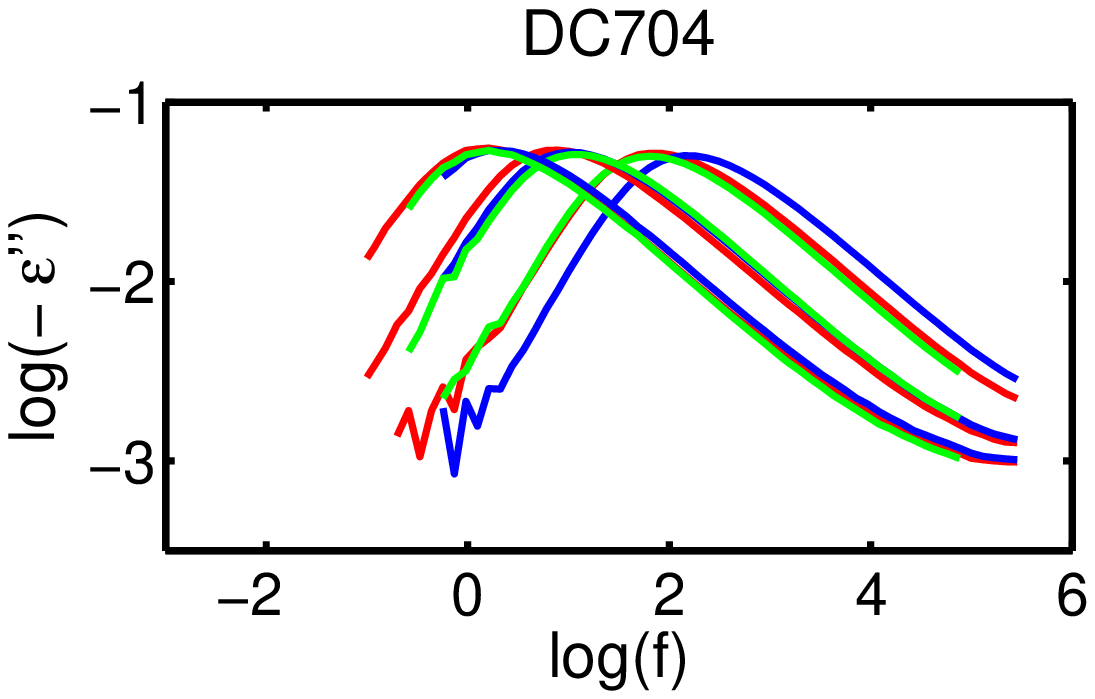}
\end{minipage}
\begin{minipage}{0.34\textwidth}
\centering
\includegraphics[width=1\textwidth]{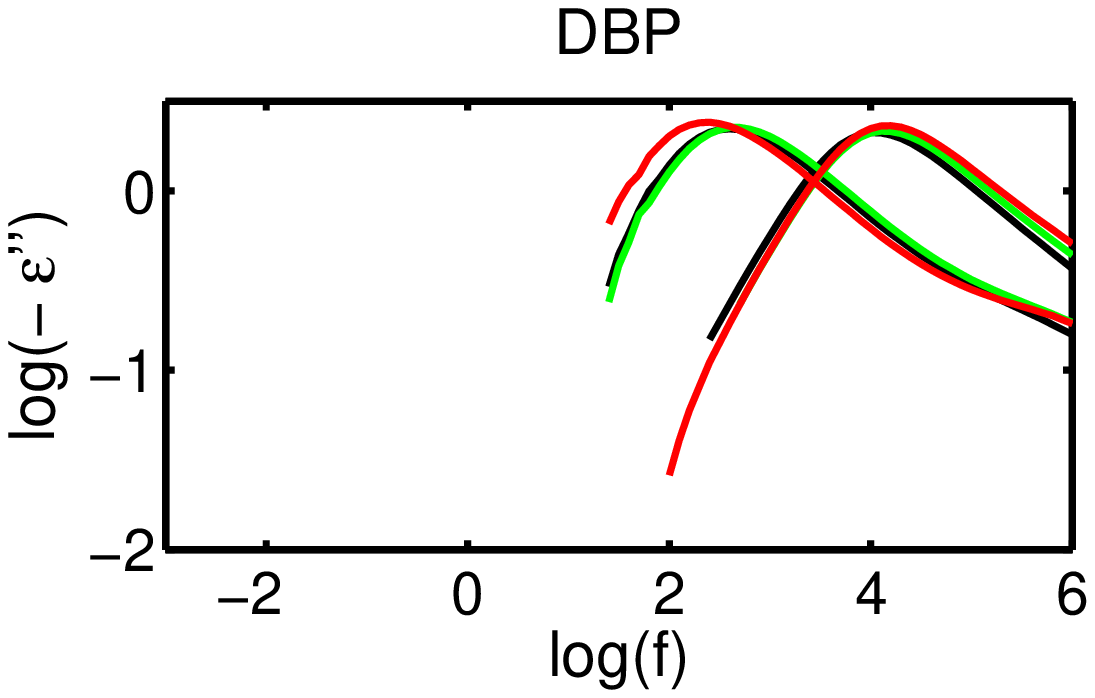}
\end{minipage}
\caption{Raw data of the six studied liquids. The
  figures show the measurements which are selected to be on
  approximate isochrones. Each color represents a pressure. The figures
  illustrate that there are differences in noise and
  signal between the liquids. Furthermore, it is seen that the
  frequency range where the measurements are performed is different
  for the different liquids.}
\label{rawdata}
\end{figure}

For each liquid, state points which are approximately on the same
isochrone are chosen for the investigation. These raw data are
shown for all 6 liquids in Fig. \ref{rawdata}. The raw data will be available at the data repository on our homepage (http://glass.ruc.dk/data) \cite{datarepository}.  We identify the
isochrones as state points with the same value loss peak frequency,
$f_{m}$, corresponding to a relaxation time $\tau=\frac{1}{2\pi f_{m}}$.
This results in 2-5 isochrones with 3-6 state points each for each
liquid. See section \ref{datas} in this supplementary for details on the selected isochrones.

\subsection{Further results}

\begin{figure}
	\centering
		\includegraphics[width=0.4\textwidth]{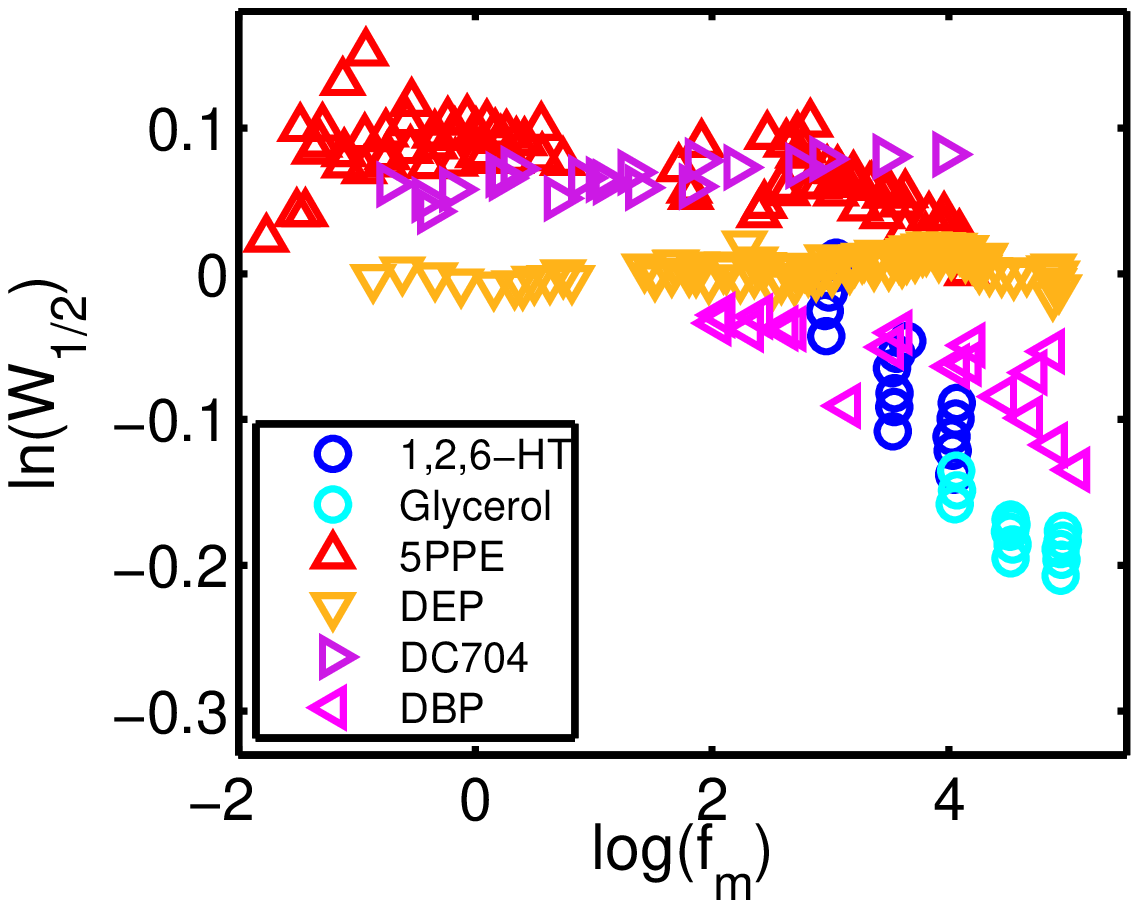}
	\caption{$W_{1/2}$ as a function of $f_{m}$ for all the measurements. 
	If the liquids obey isochronal superposition $W_{1/2}$ has to be the 
	same when $f_{m}$ is the same.}
	\label{figure2s}
\end{figure}

The data which are used in the article are selected so that they are approximately on isochrones. However, we have performed more measurements. In Fig. \ref{figure2s}, $W_{1/2}$ for all the data is shown as a
function of $f_{m}$. It is seen that the $W_{1/2}$-values for
especially DEP and DC704 are close to being the identical when $f_{m}$
is the same. This is also the case for 5PPE and DBP at some values of
$f_{m}$, but not all. For glycerol and 1,2,6-HT on the other hand,
$W_{1/2}$ seems to vary for all values of $f_{m}$. If the liquid obeys
isochronal superposition, then $W_{1/2}$ has to be the same when
$f_{m}$ is the same. Thus we can see directly from this figure that
glycerol and 1,2,6-HT deviate from IS.

\begin{figure}
	\centering
		\includegraphics[width=0.4\textwidth]{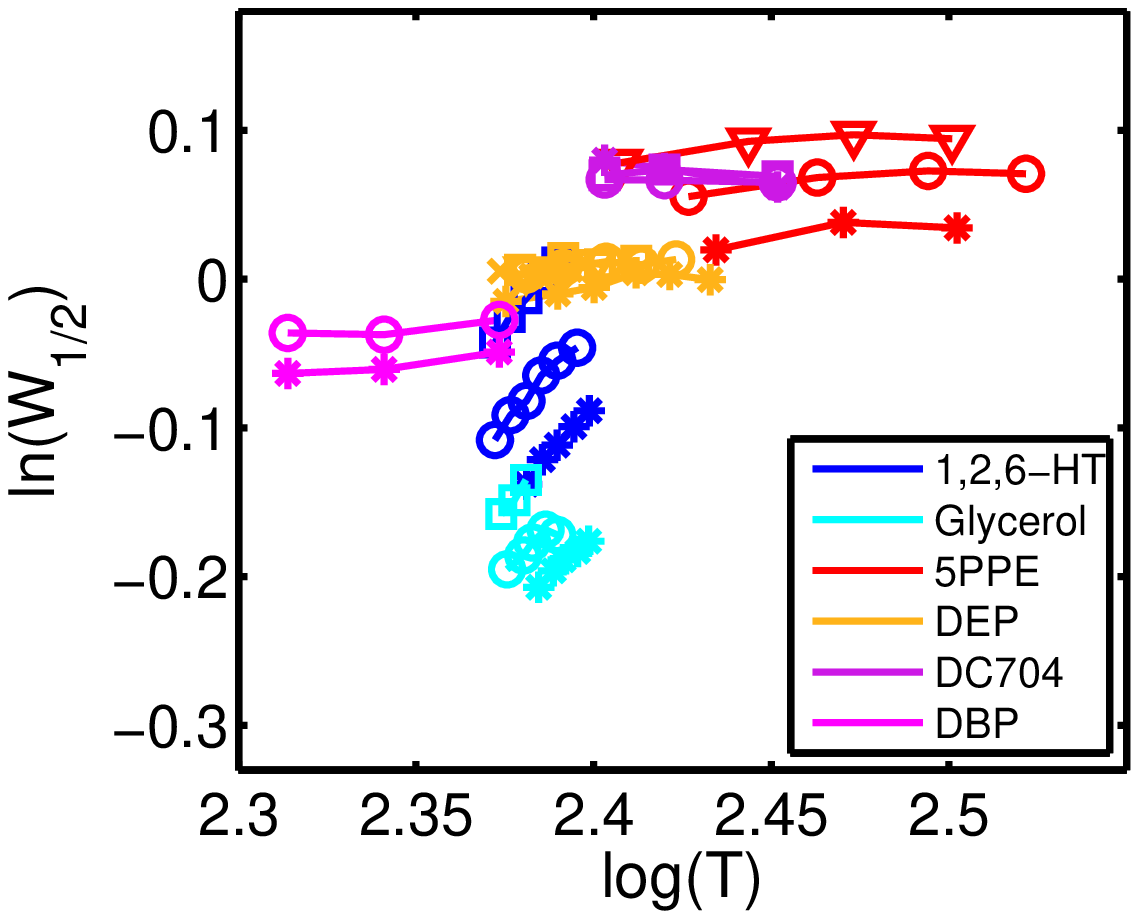}
                \caption{$W_{1/2}$ as a function of temperature for
                  each isochrone. Blue: 1,2,6-HT, stars: log$(f_{m})\simeq$
                  4.0, circles: log$(f_{m})\simeq$ 3.6, squares:
                  log$(f_{m})\simeq$ 3.0. Cyan: glycerol, stars:
                  log$(f_{m})\simeq$5.0, circles: log$(f_{m})\simeq$
                  4.5, squares: log$(f_{m})\simeq$ 4.1. Red: 5PPE,
                  stars: log$(f_{m})\simeq$ 4.0, circles: log$(f_{m})\simeq$
                  3.1, triangles: log$(f_{m})\simeq0.2$. Orange: DEP, stars: 
                  log$(f_{m})\simeq4.9$, circles: log$(f_{m})\simeq4.3$, squares:
                  log$(f_{m})\simeq3.4$, triangles: log$(f_{m})\simeq2.4$, 
                  crosses: log$(f_{m})\simeq1.7$. Purple: DC704, stars: 
                  log$(f_{m})\simeq2.0$, circles: log$(f_{m})\simeq1.0$, squares: 
                  log$(f_{m})\simeq0.3$. Magenta: DBP, stars log$(f_{m})\simeq4.1$,
                  circles: log$(f_{m})\simeq2.6$.}
	\label{figure3s}
\end{figure}

Figure \ref{figure3s} is the same figure as Fig. 2(a) in the article. Here, the caption is more informative. The exact state points can be found in section \ref{datas} in this supplementary.

\subsection{Calculation of the area of the spectrum}

 As one of the shape parameters, we use the area of the relaxation spectrum ($A$). $A$ is found by adding data taken at -0.4, -0.2, 0.2, 0.4, 0.6, 0.8, and 1.0 decades relative to the loss peak frequency, 
 $|\sum^{7}_{i=1}\text{log}(\tilde{\varepsilon}_{j+1}(\tilde{f}_i))|$, where $\tilde{\varepsilon}=\frac{\varepsilon''}{\varepsilon_{max}''}$ and
$\tilde{f}=\frac{f}{f_{m}}$. 

\subsection{Cole-Davidson fit}

As a model-dependent shape parameter, we use $\beta_{CD}$ from the Cole-Davidson fitting function \cite{cole1950}
\begin{equation}
 \tilde{\varepsilon}(\omega)=\varepsilon_{\infty}+\frac{\Delta\varepsilon}{(1+i\omega\tau)^{\beta_{CD}}}
\end{equation}

\begin{figure}
\begin{minipage}{0.3\textwidth}
\centering
\includegraphics[width=1\textwidth]{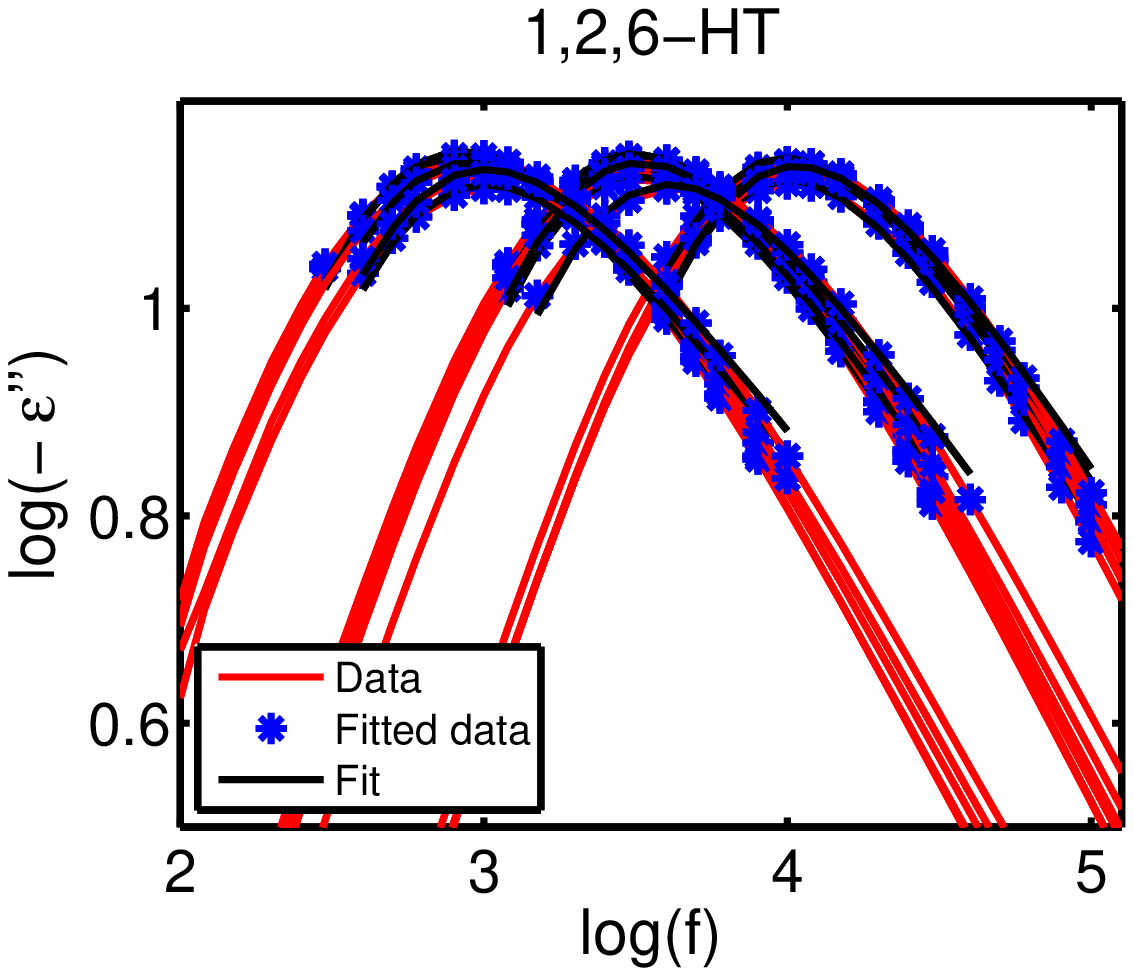}
\end{minipage}
\begin{minipage}{0.3\textwidth}
\centering
\includegraphics[width=1\textwidth]{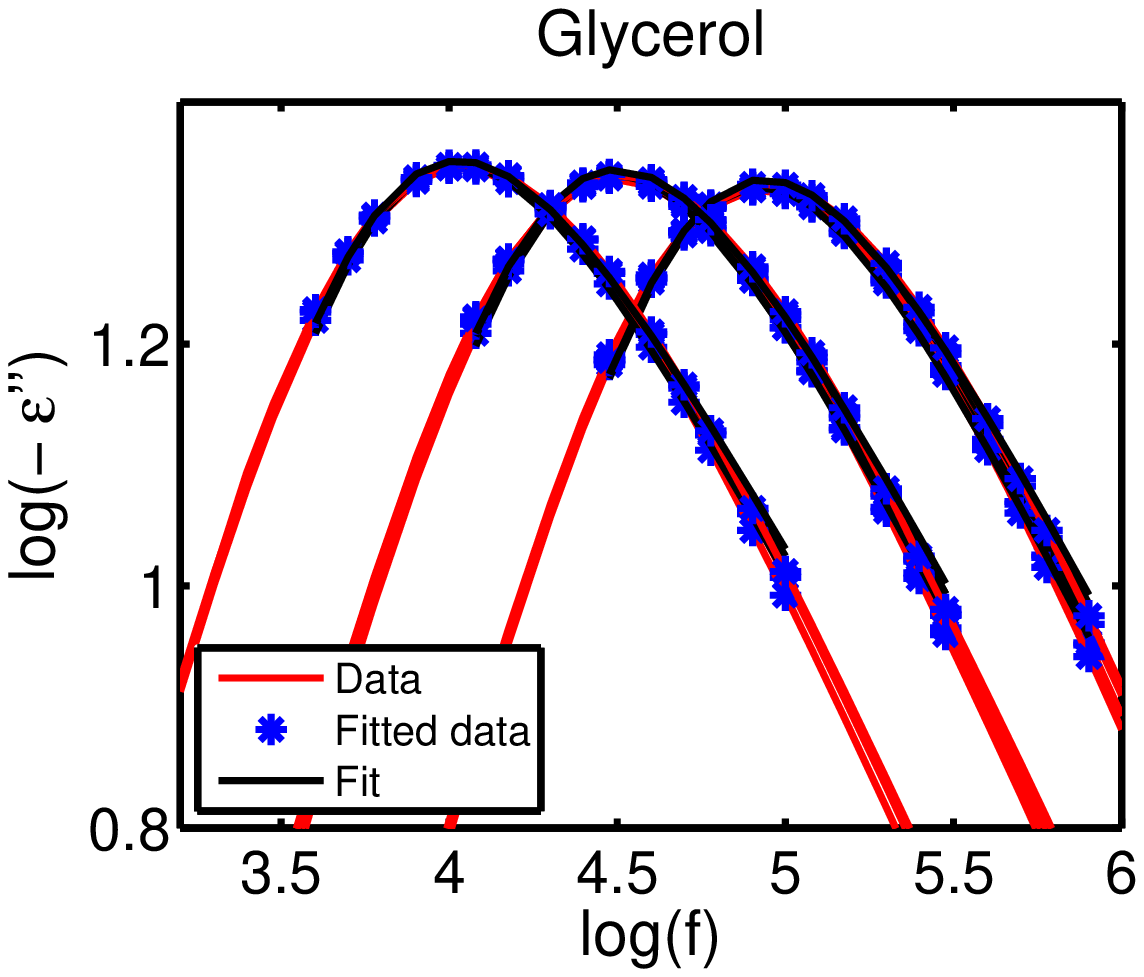}
\end{minipage}
\begin{minipage}{0.3\textwidth}
\centering
\includegraphics[width=1\textwidth]{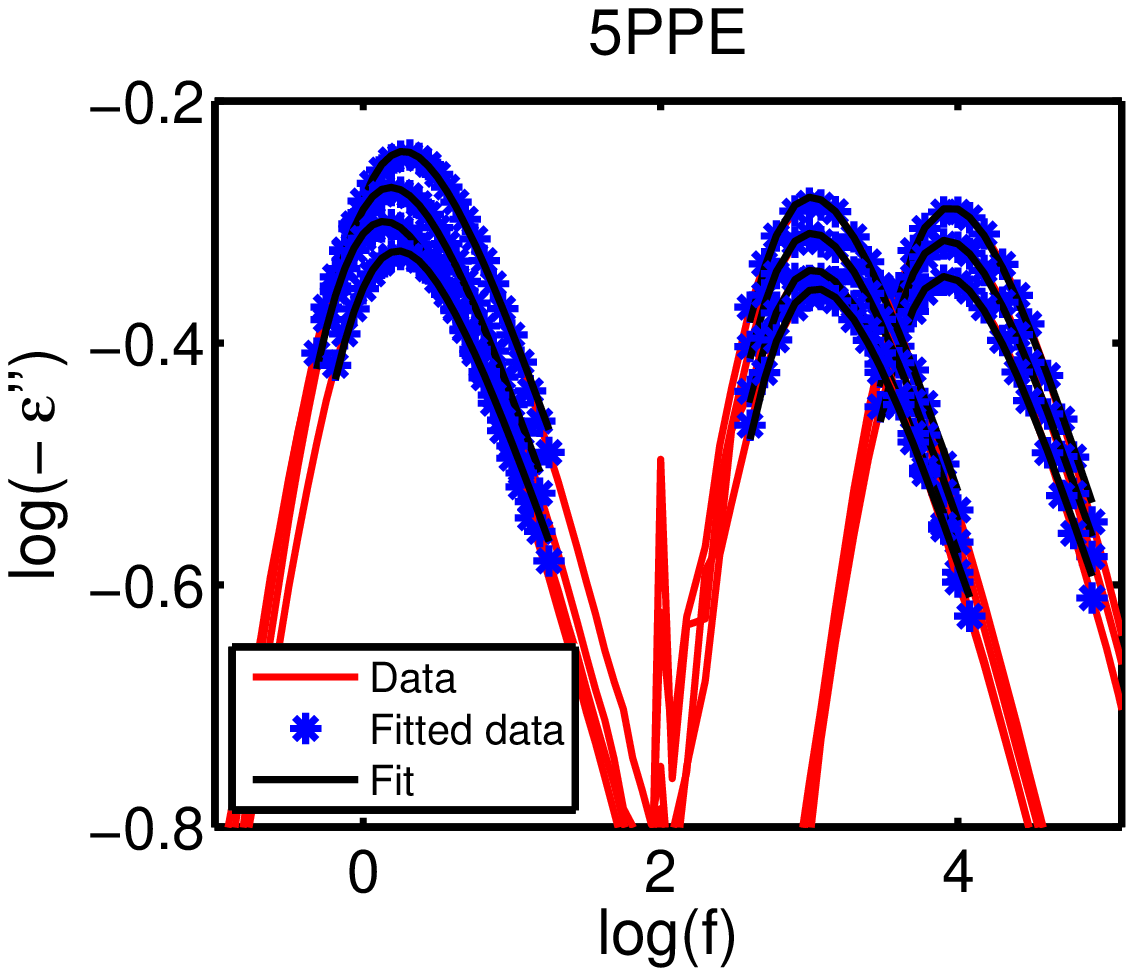}
\end{minipage}
\begin{minipage}{0.3\textwidth}
\centering
\includegraphics[width=1\textwidth]{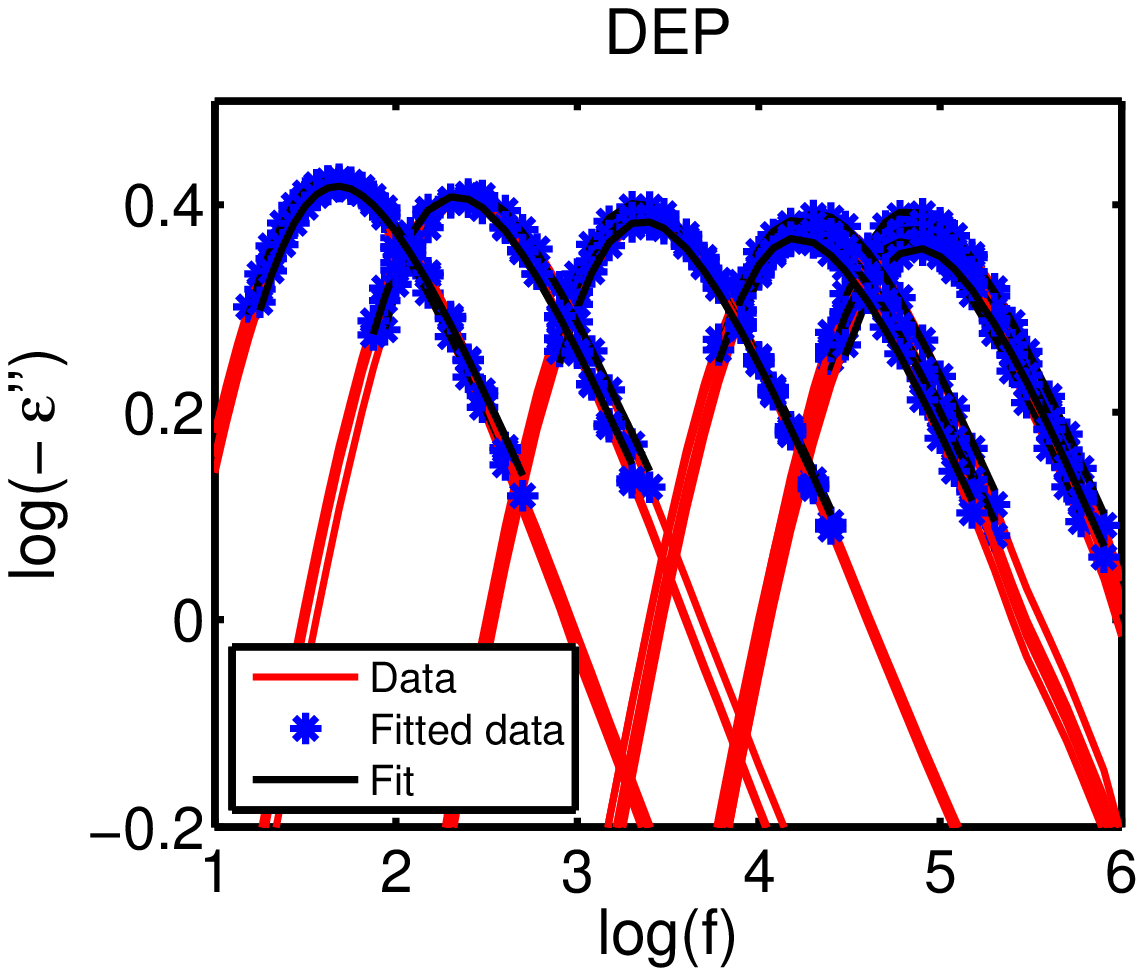}
\end{minipage}
\begin{minipage}{0.3\textwidth}
\centering
\includegraphics[width=1\textwidth]{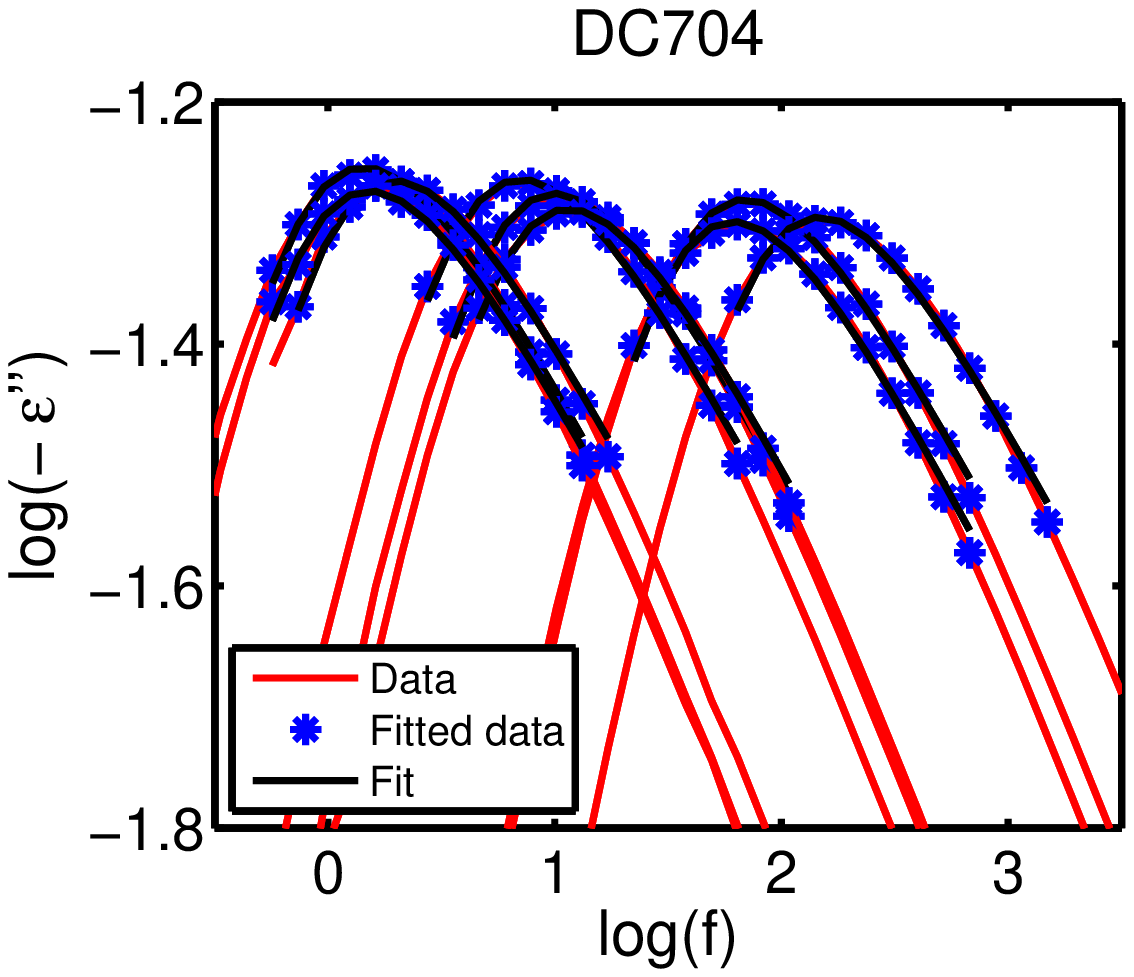}
\end{minipage}
\begin{minipage}{0.3\textwidth}
\centering
\includegraphics[width=1\textwidth]{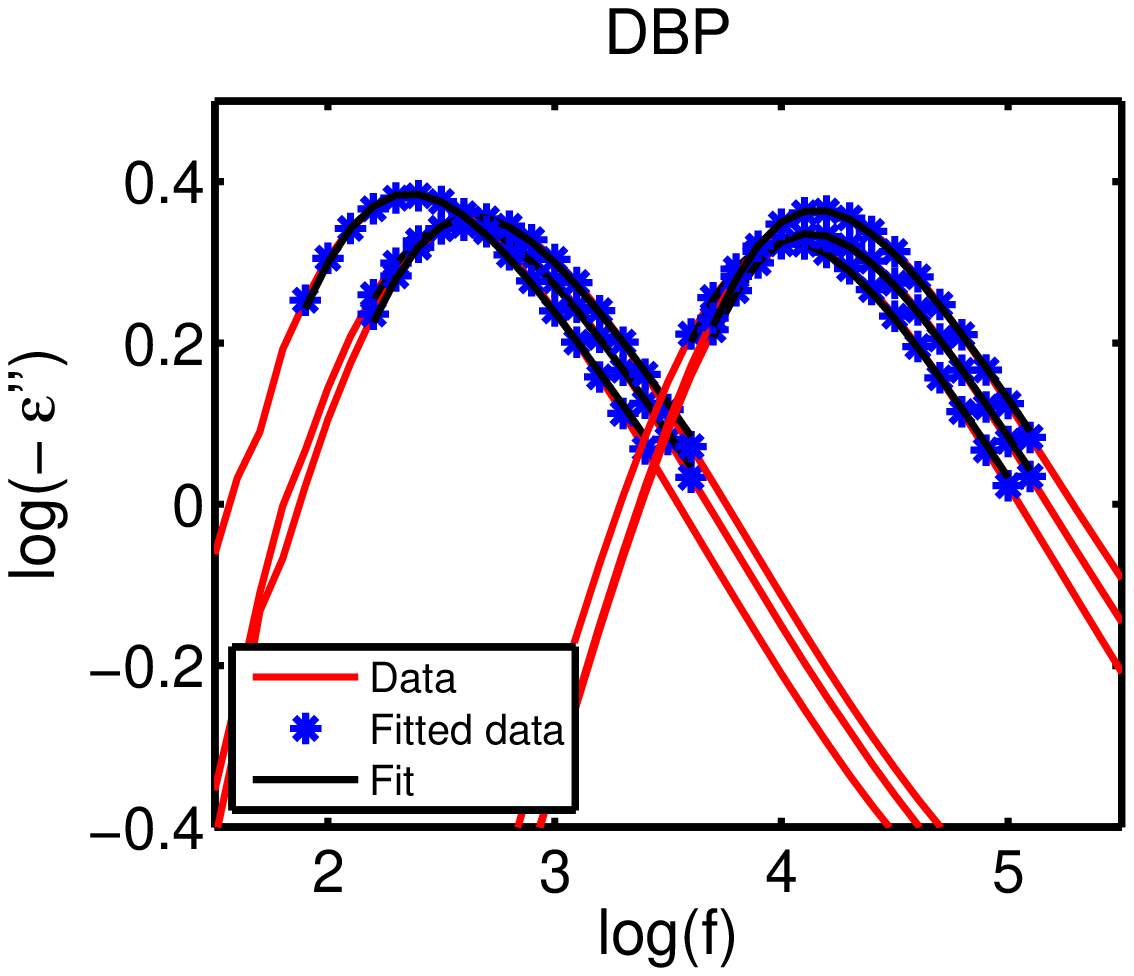}
\end{minipage}
\caption{Here the raw data, the part of the raw data which is used for the fit and the Cole-Davidson fit are seen.}
\label{figfit}
\end{figure}

For all spectra we have fitted from half a decade on the low-frequency side of $f_{m}$ to one decade on the high-frequency side of $f_{m}$. This interval is chosen, so that all spectra are fitted in the same size of interval. The fits is shown in Fig. \ref{figfit}.

\subsection{Uncertainty in the measures}\label{error}

We have tried to find the experimental uncertainty in the measurements to estimate an error bar for the measures. However, there are many different ways to estimate the error bars, and this is just one way to do it.

We find the error bars for the shape parameter ($X$) using the relative deviation of the shape parameter of two measurements at the same state point. We then fit a power law to the maximum relative deviation for each liquid against the dielectric strength of the liquid. The maximum deviations are used, because the measurements at the same state point are made right after each other and we expect that the deviation would be larger, if the two measurements where totally independent.

The fitted power law is then used to find the relative deviation $p$ for each liquid, which is then used to calculate the error bar of the shape parameter $\sigma_{X}$ for each liquid in this way
\begin{equation}
 \sigma_{X}=X\cdot p
\end{equation}
where $p$ is the relative deviation found by the fitted power law.

We then find the error bar of $\ln(X)$ ($\sigma_{\ln(X)}$) in the following way
\begin{equation}
\sigma_{\ln(X)}=\ln(X)-\ln(X\cdot(1-p))=\ln\left(\frac{1}{1-p}\right)
\end{equation}
The error bar of the measure $L(X)$ ($\sigma_{L(X)}$) is then found as
\begin{equation}
 \sigma_{L(X_{i+1/2})}=\frac{\sqrt{2}\cdot\sigma_{\ln(X)}}{\sqrt{\ln^2(T_i/T_{i+1})+\ln^2(\rho_i/\rho_{i+1})}}
\end{equation}
where we use a standard method for calculating the error bar of a sum of (or difference between) two numbers. We assume that there is no uncertainty in the temperature and density.

The error bar of the measure $L_T(X)$ ($\sigma_{L_T(X)}$) is found as follows
\begin{equation}
 \sigma_{L_T(X_{i+1/2})}=\frac{\sqrt{2}\cdot\sigma_{\ln(X)}}{\ln(T_i/T_{i+1})}
\end{equation}

\begin{figure}
\begin{minipage}{0.40\textwidth}
\centering
\includegraphics[width=1\textwidth]{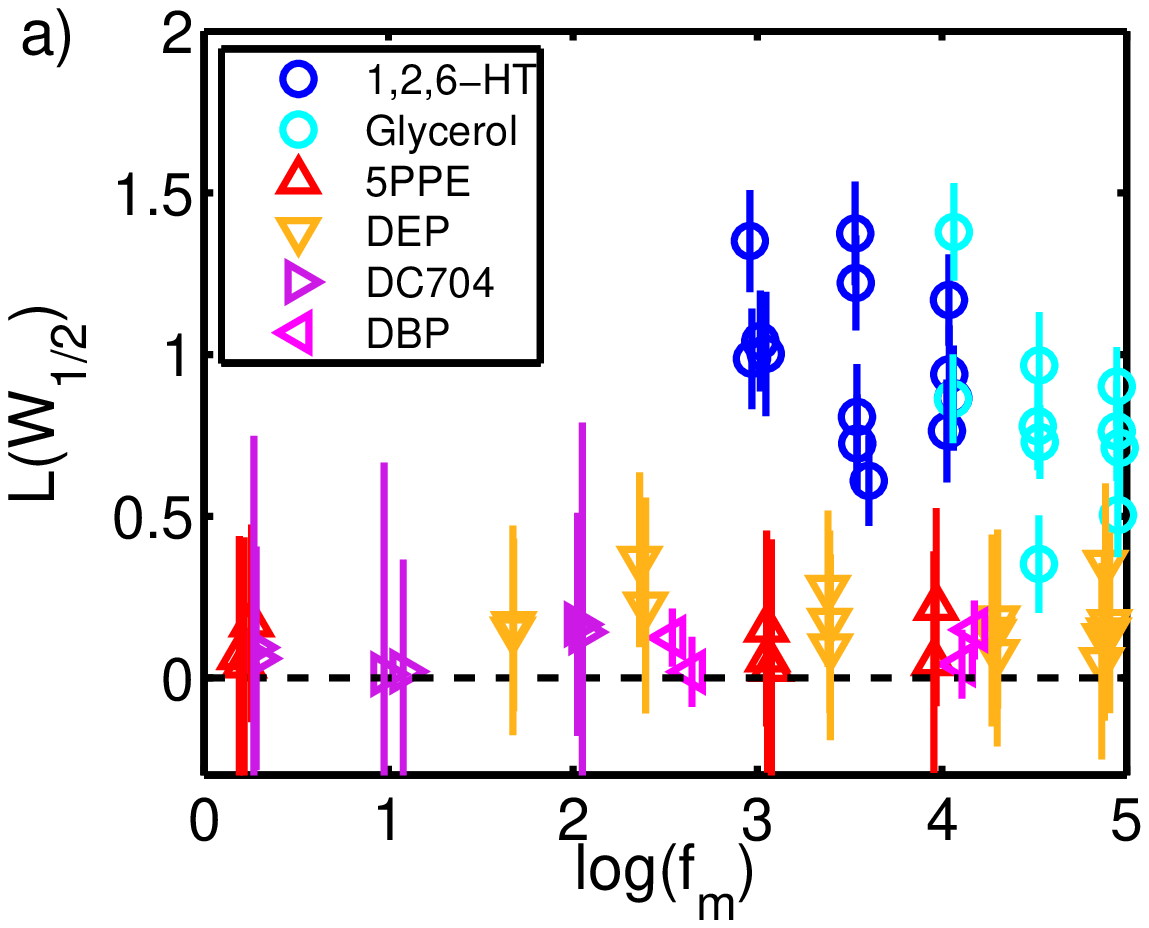}
\end{minipage}
\begin{minipage}{0.40\textwidth}
\centering
\includegraphics[width=1\textwidth]{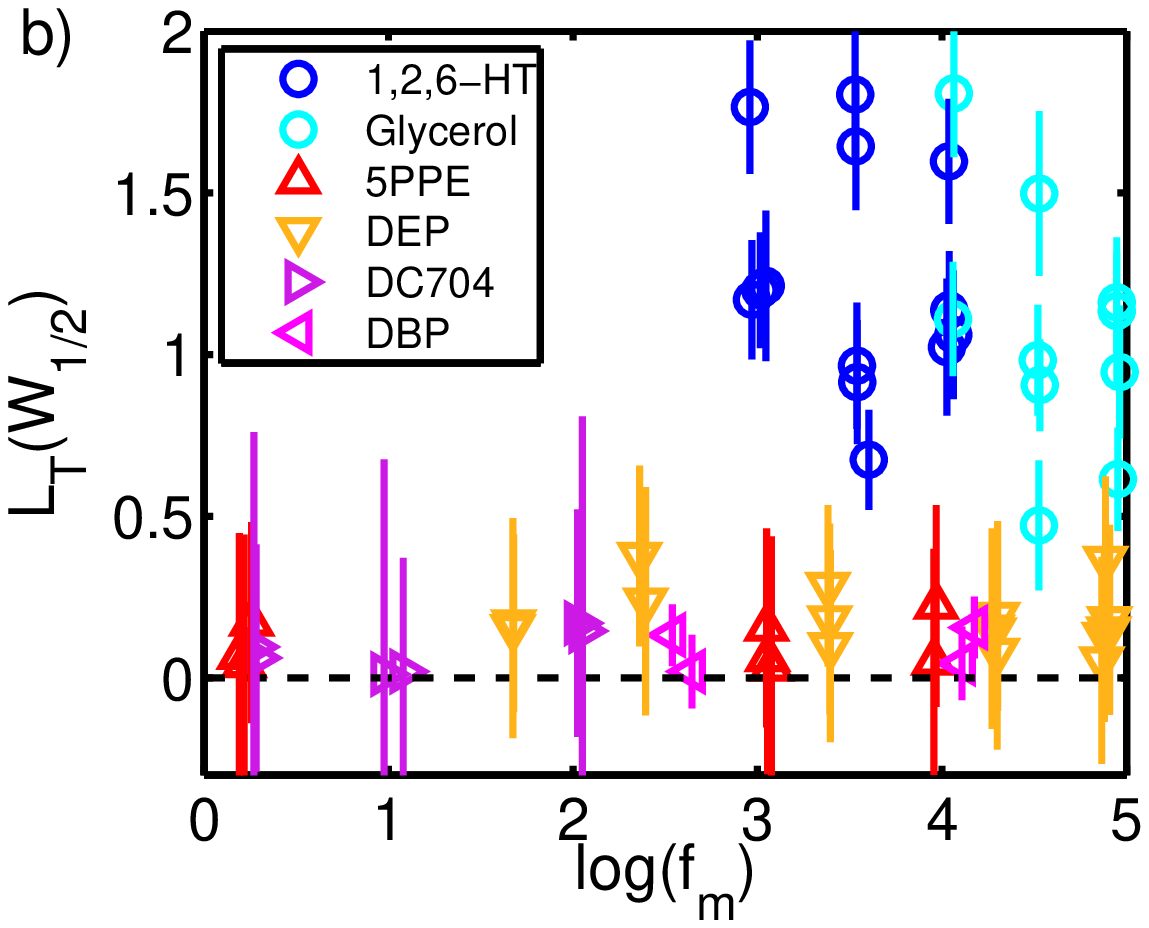}
\end{minipage}
\caption{The measures based on $W_{1/2}$ with the error bars calculated as described in section \ref{error} in this supplementary.
(a) The measure $L(W_{1/2})$. 
(b) The measure $L_T(W_{1/2})$. 
If the liquid obeys IS, both measures must be zero within the experimental uncertainty. This is the case for the four van der Waals liquids, but not for the two hydrogen-bonded liquids. }
\label{figure4s}
\end{figure}

\begin{figure}
\begin{minipage}{0.40\textwidth}
\centering
\includegraphics[width=1\textwidth]{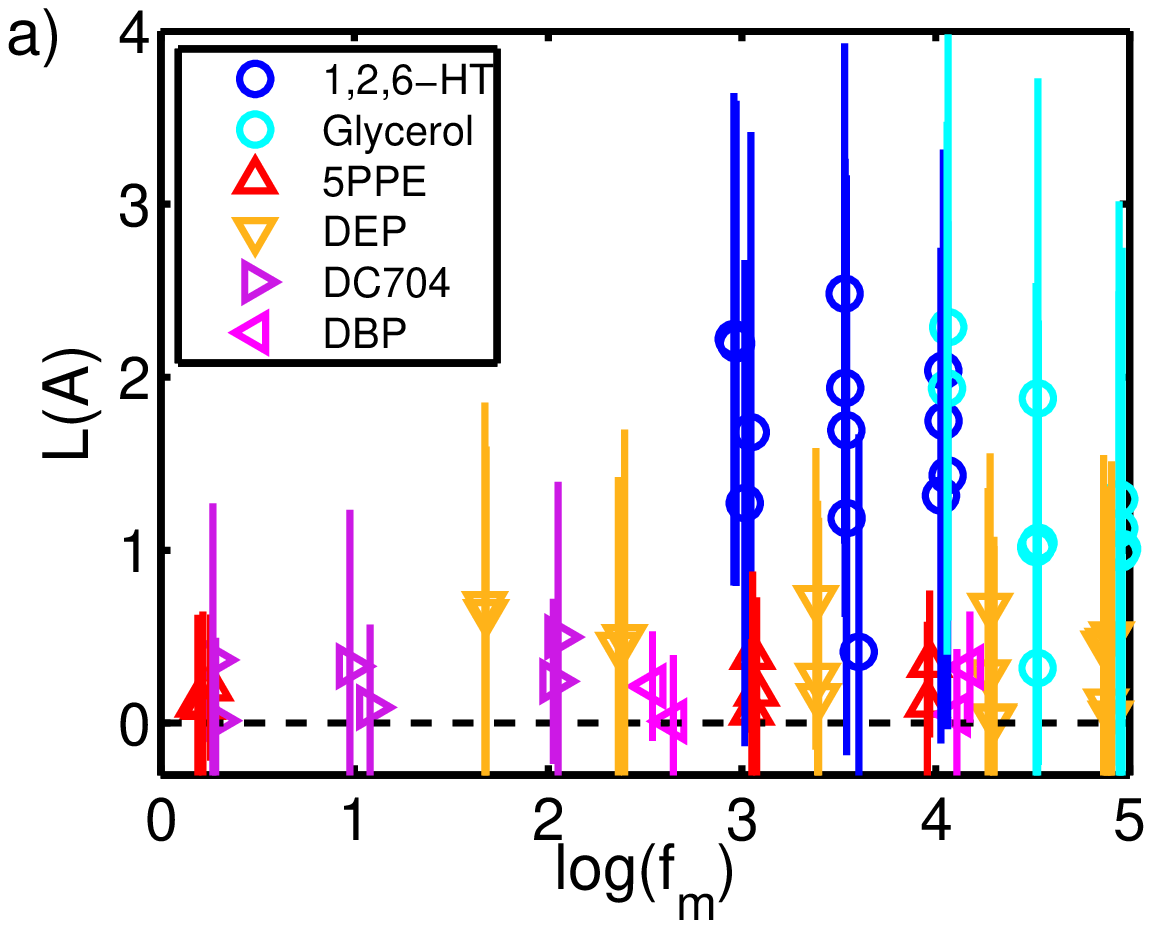}
\end{minipage}
\begin{minipage}{0.40\textwidth}
\centering
\includegraphics[width=1\textwidth]{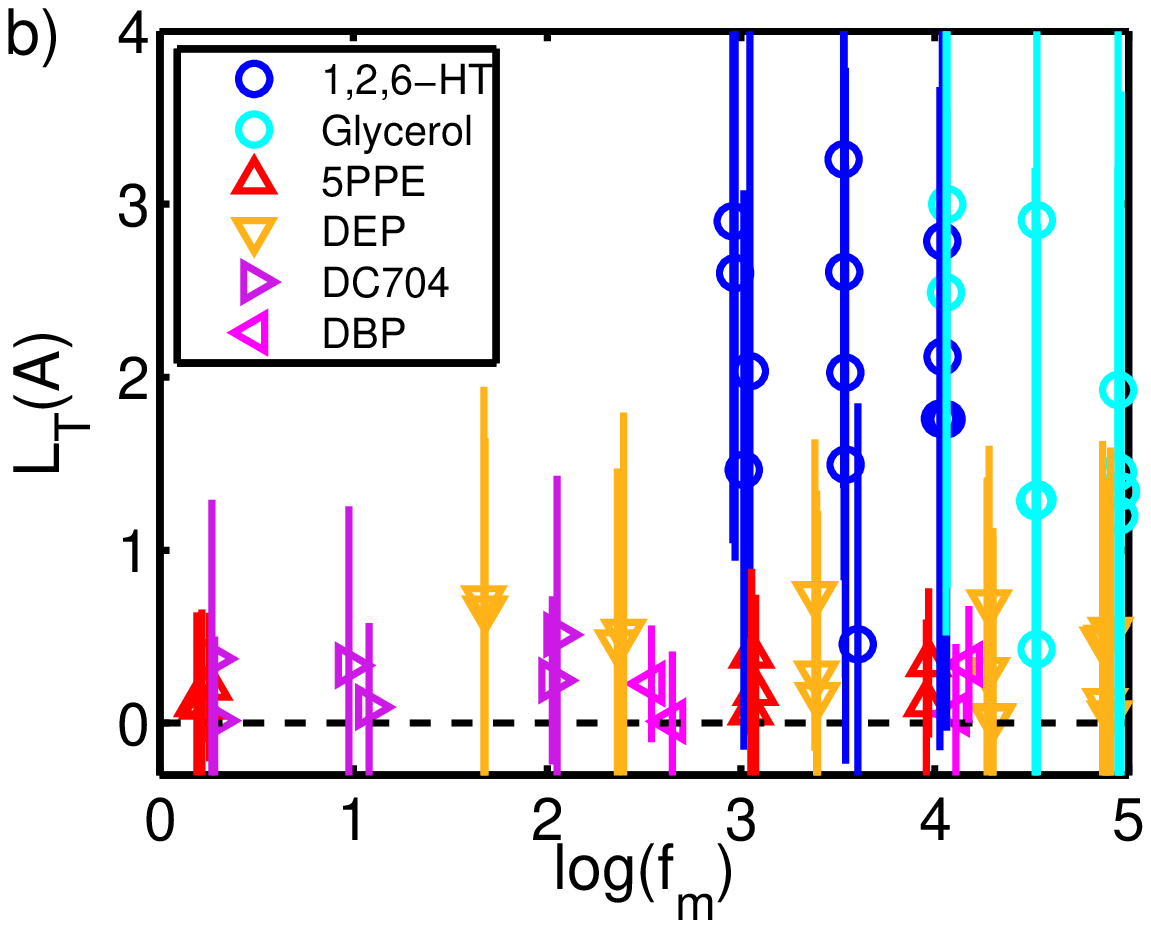}
\end{minipage}
\caption{The measures based on $A$ with the error bars calculated as described in section \ref{error} in this supplementary. 
(a) The measure $L(A)$.
(b) The measure $L_T(A)$.}
\label{figure5s}
\end{figure}

The Figs. \ref{figure4s} and \ref{figure5s} show the measures with these error bars. However, there is many different ways of estimating and calculating error bars.

\subsection{Qualitative analysis of isochronal superposition}\label{isttps}

A qualitative picture of whether the liquids obey isochronal superposition or not is obtained by normalizing the dielectric relaxation spectra, such that they have the same maximum, which is also what Roland \textit{et al}.\cite{roland2003} and Ngai \textit{et al}.\cite{ngai2005} have done in their data treatment with state points that almost are on the same isochrone. However, we normalize all the used dielectric relaxation spectra for a liquid, such that it is also seen whether the liquids obey isochronal superposition better than time-temperature-pressure superposition (TTPS). This type of data treatment is also used in Ref. \onlinecite{niss2007}.

\begin{figure}
\begin{minipage}{0.3\textwidth}
\centering
\includegraphics[width=1\textwidth]{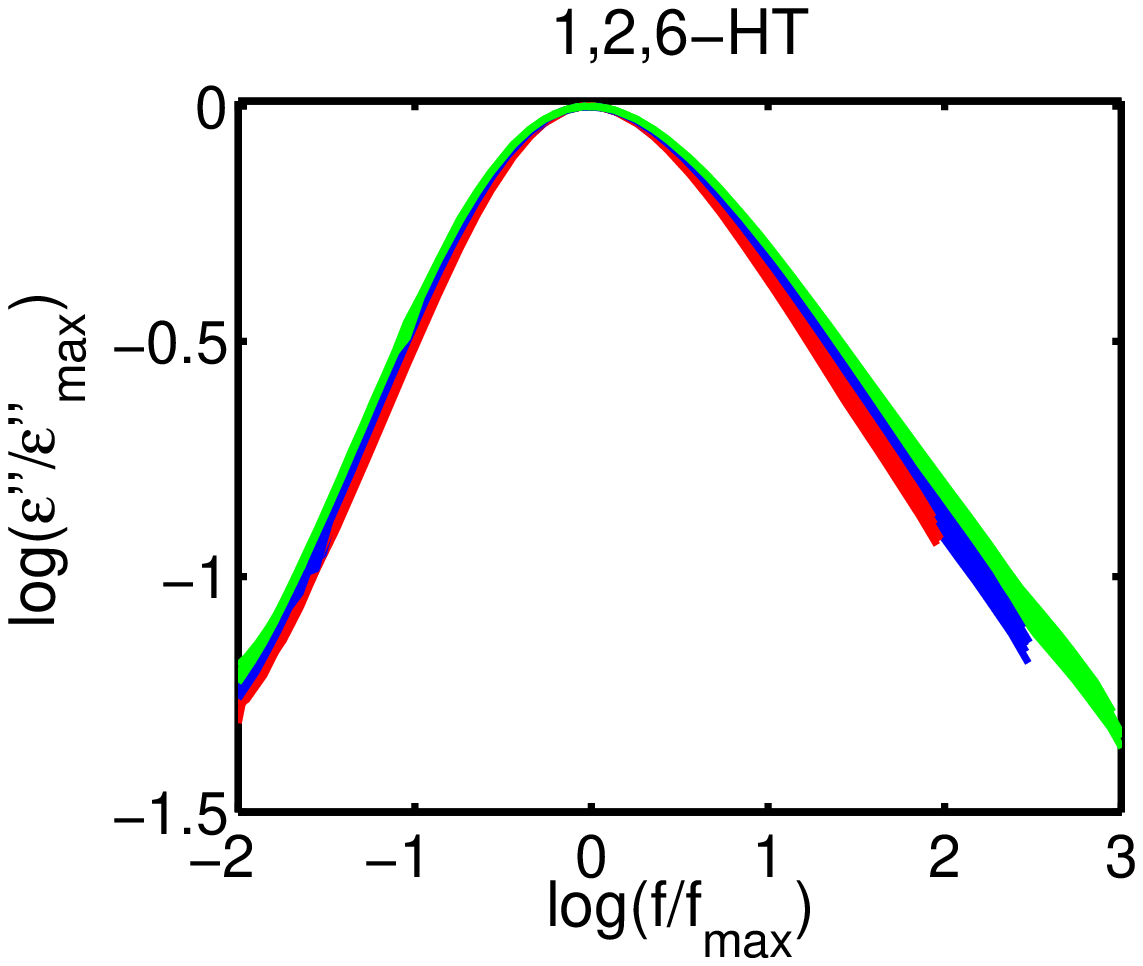}
\end{minipage}
\begin{minipage}{0.3\textwidth}
\centering
\includegraphics[width=1\textwidth]{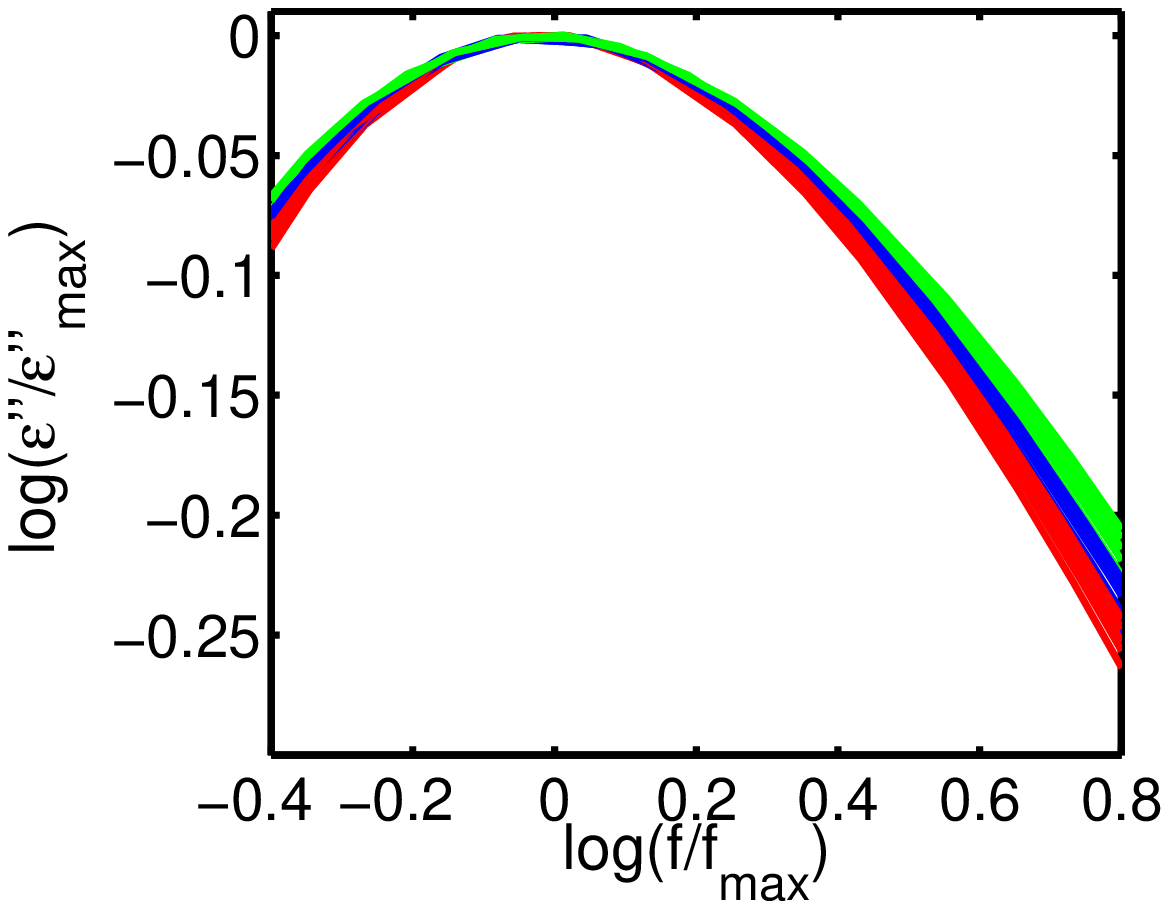}
\end{minipage}
\caption{The normalized relaxation spectra for 1,2,6-HT. The three different colors show the three isochrones. The figure to the right is a zoom of the peak. The red spectra are measurements with log$(f_{m})\simeq$ 4, the blue spectra are measurements with log$(f_{m})\simeq$ 3.6, and the green spectra are measurements with log$(f_{m})\simeq$ 3.}
\label{126HTTTPS}
\end{figure}

\begin{figure}
\begin{minipage}{0.3\textwidth}
\centering
\includegraphics[width=1\textwidth]{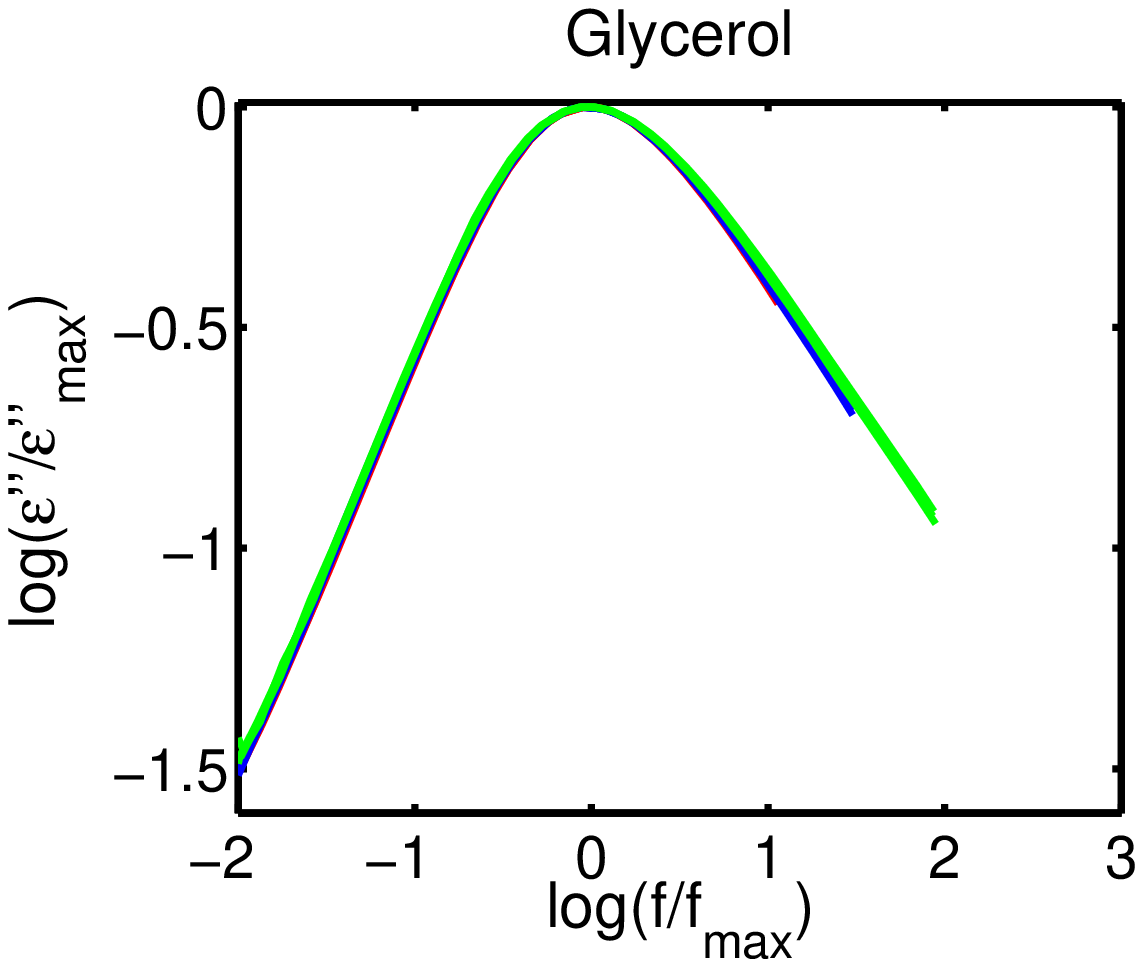}
\end{minipage}
\begin{minipage}{0.3\textwidth}
\centering
\includegraphics[width=1\textwidth]{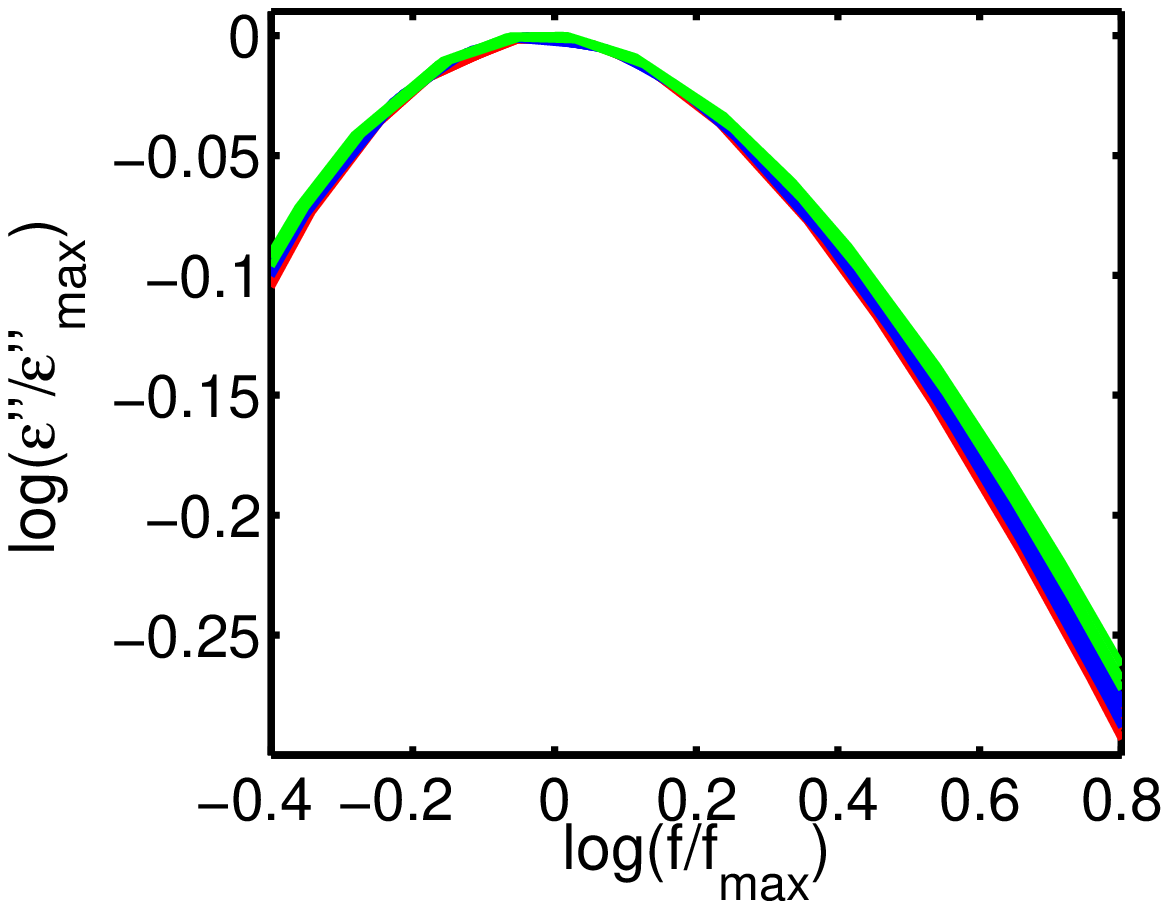}
\end{minipage}
\caption{The normalized relaxation spectra for glycerol. The three different colors show the three isochrones. The figure to the right is a zoom of the peak. The red spectra are measurements with log$(f_{m})\simeq$ 5, the blue spectra are measurements with log$(f_{m})\simeq$ 4.5, and the green spectra are measurements with log$(f_{m})\simeq$ 4.}
\label{glycerolTTPS}
\end{figure}

\begin{figure}
\begin{minipage}{0.3\textwidth}
\centering
\includegraphics[width=1\textwidth]{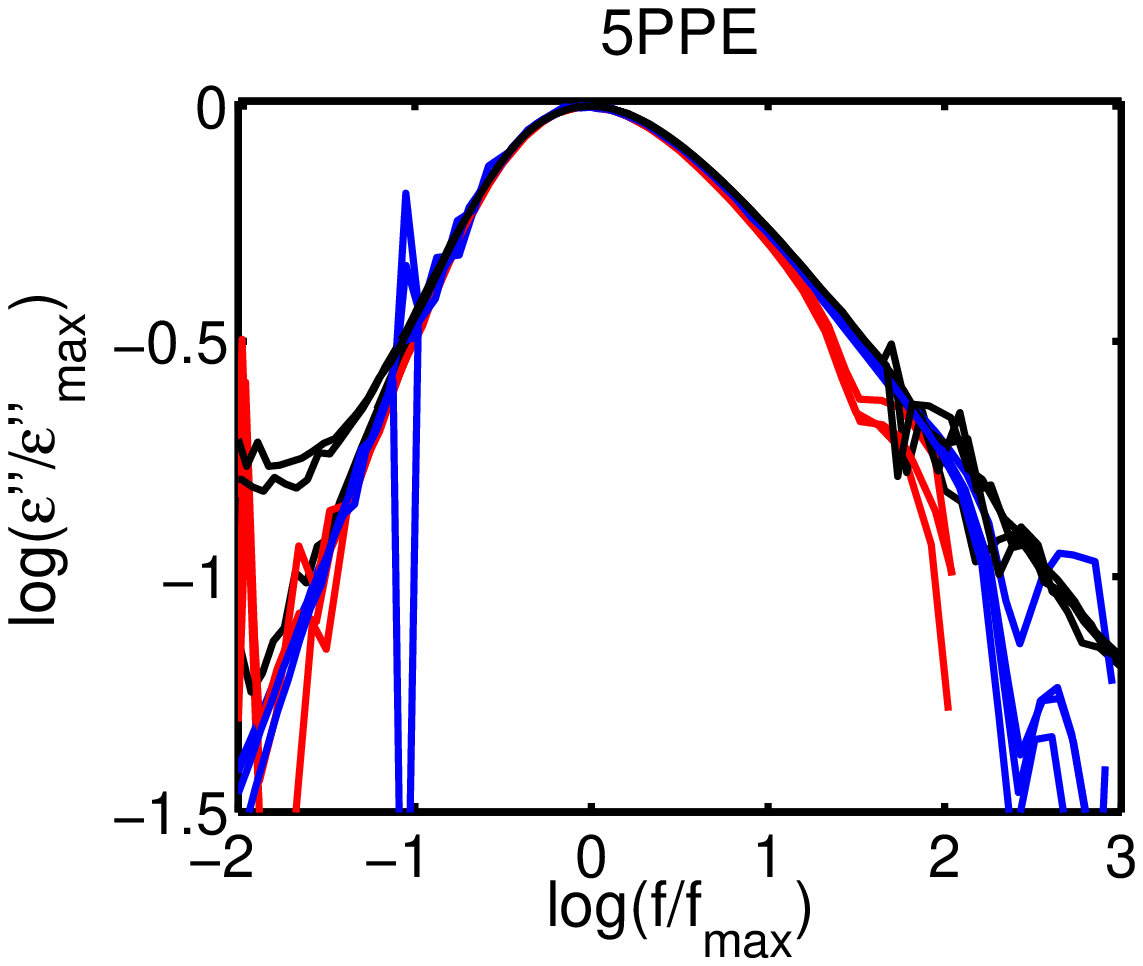}
\end{minipage}
\begin{minipage}{0.3\textwidth}
\centering
\includegraphics[width=1\textwidth]{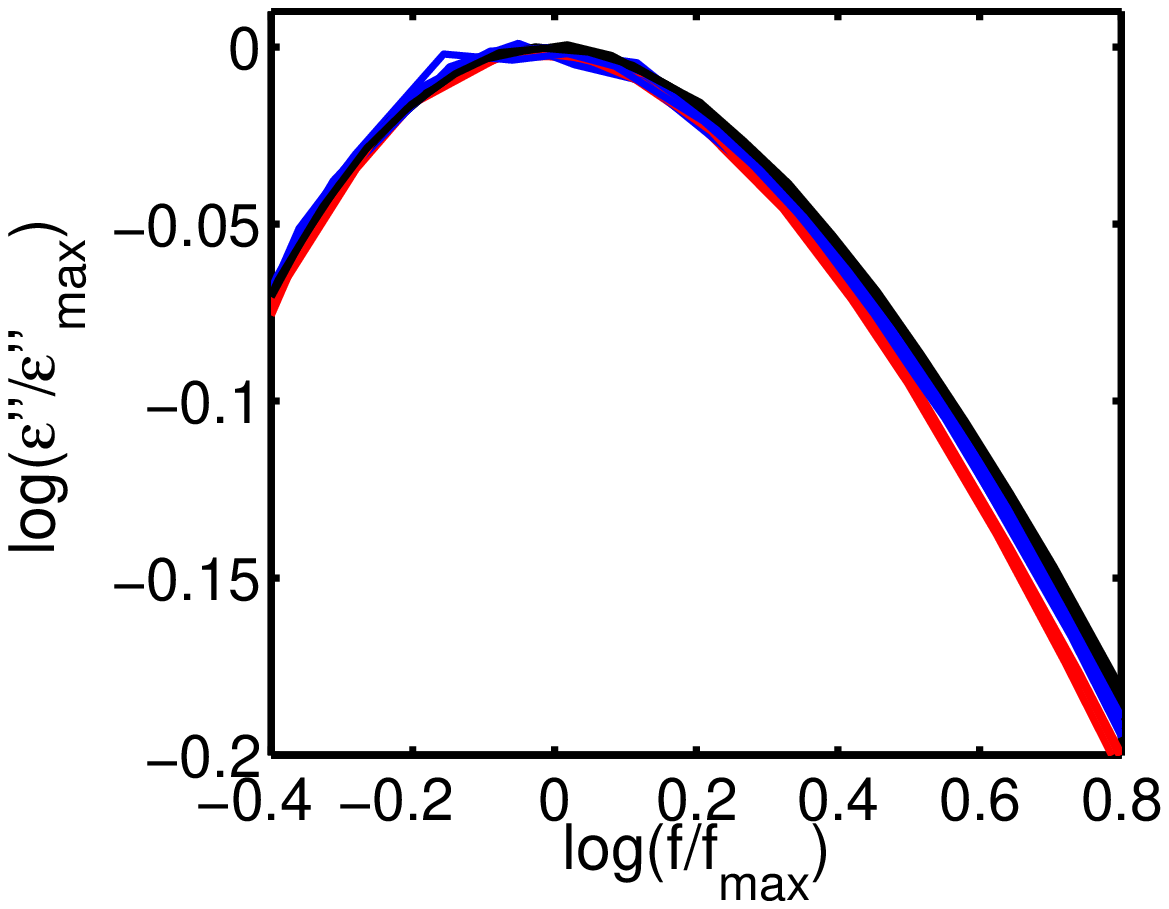}
\end{minipage}
\caption{The normalized relaxation spectra for 5PPE. The three different colors show the three isochrones. The figure to the right is a zoom of the peak. The red spectra are measurements with log$(f_{m})\simeq$ 4, the blue spectra are measurements with log$(f_{m})\simeq$ 3.1, and the black spectra are measurements with log$(f_{m})\simeq$ 0.2.}
\label{5PPETTPS}
\end{figure}

\begin{figure}
\begin{minipage}{0.3\textwidth}
\centering
\includegraphics[width=1\textwidth]{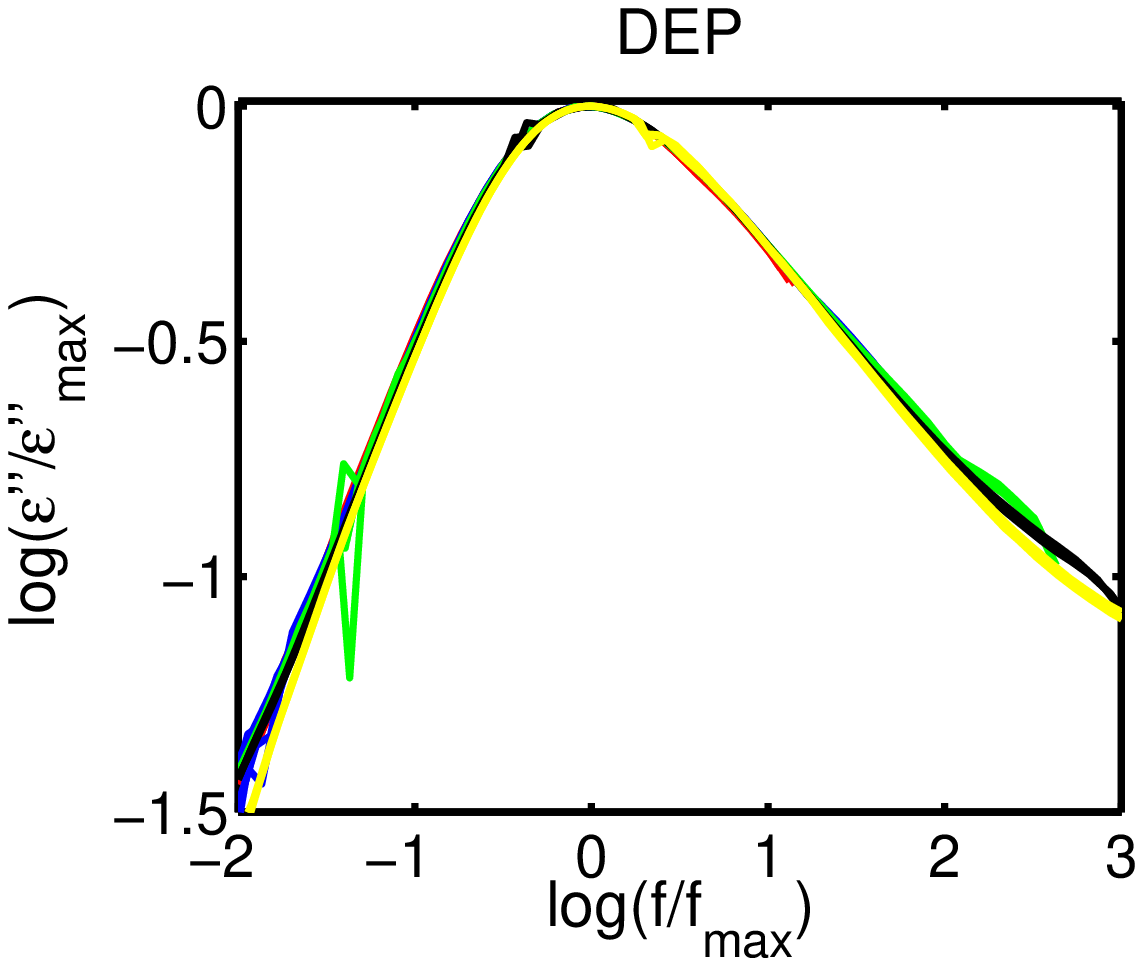}
\end{minipage}
\begin{minipage}{0.3\textwidth}
\centering
\includegraphics[width=1\textwidth]{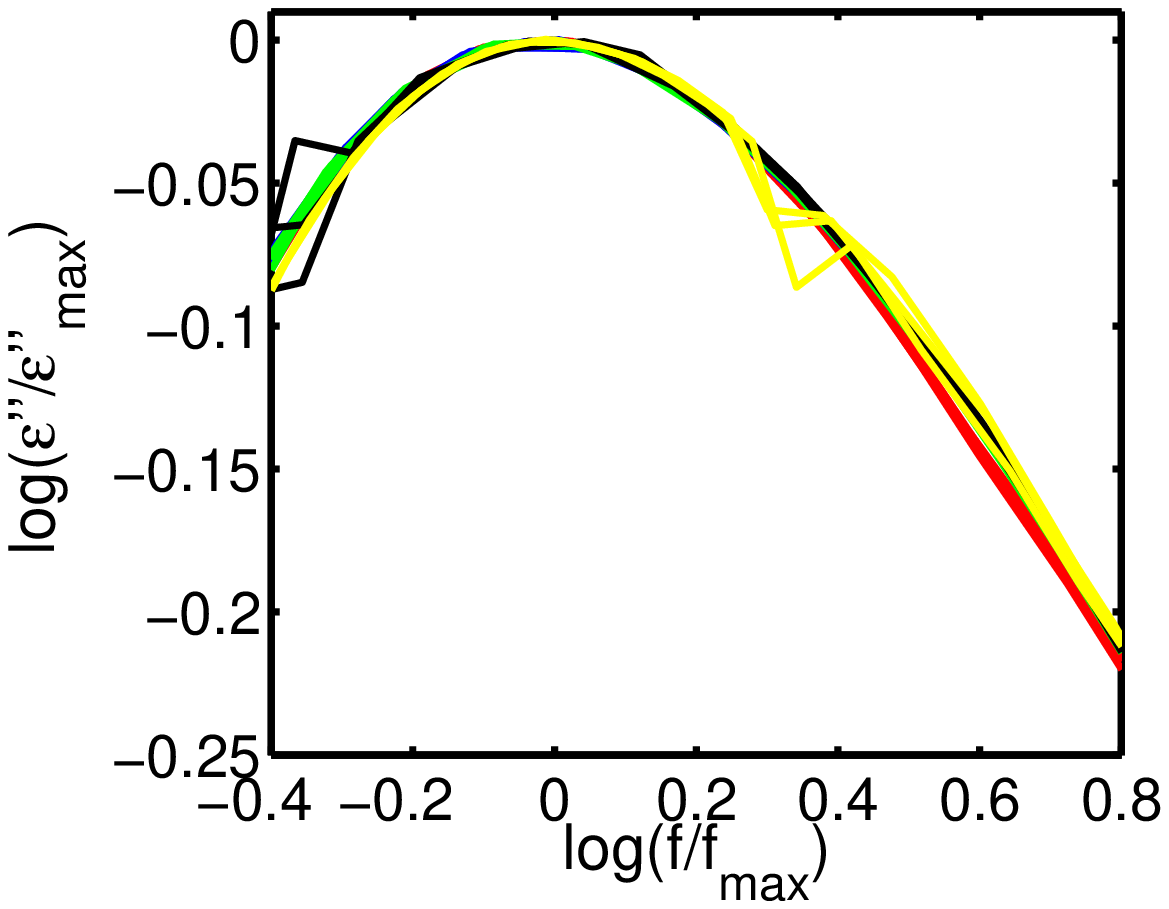}
\end{minipage}
\caption{The normalized relaxation spectra for DEP. The five different colors show the five isochrones. The figure to the right is a zoom of the peak. The red spectra are measurements with log$(f_{m})\simeq$ 4.9, the blue spectra are measurements with log$(f_{m})\simeq$ 4.3, the green spectra are measurements with log$(f_{m})\simeq$ 3.4, the black spectra are measurements with log$(f_{m})\simeq$ 2.4, and the yellow spectra are measurements with log$(f_{m})\simeq$ 1.7.}
\label{DEPTTPS}
\end{figure}

\begin{figure}
\begin{minipage}{0.3\textwidth}
\centering
\includegraphics[width=1\textwidth]{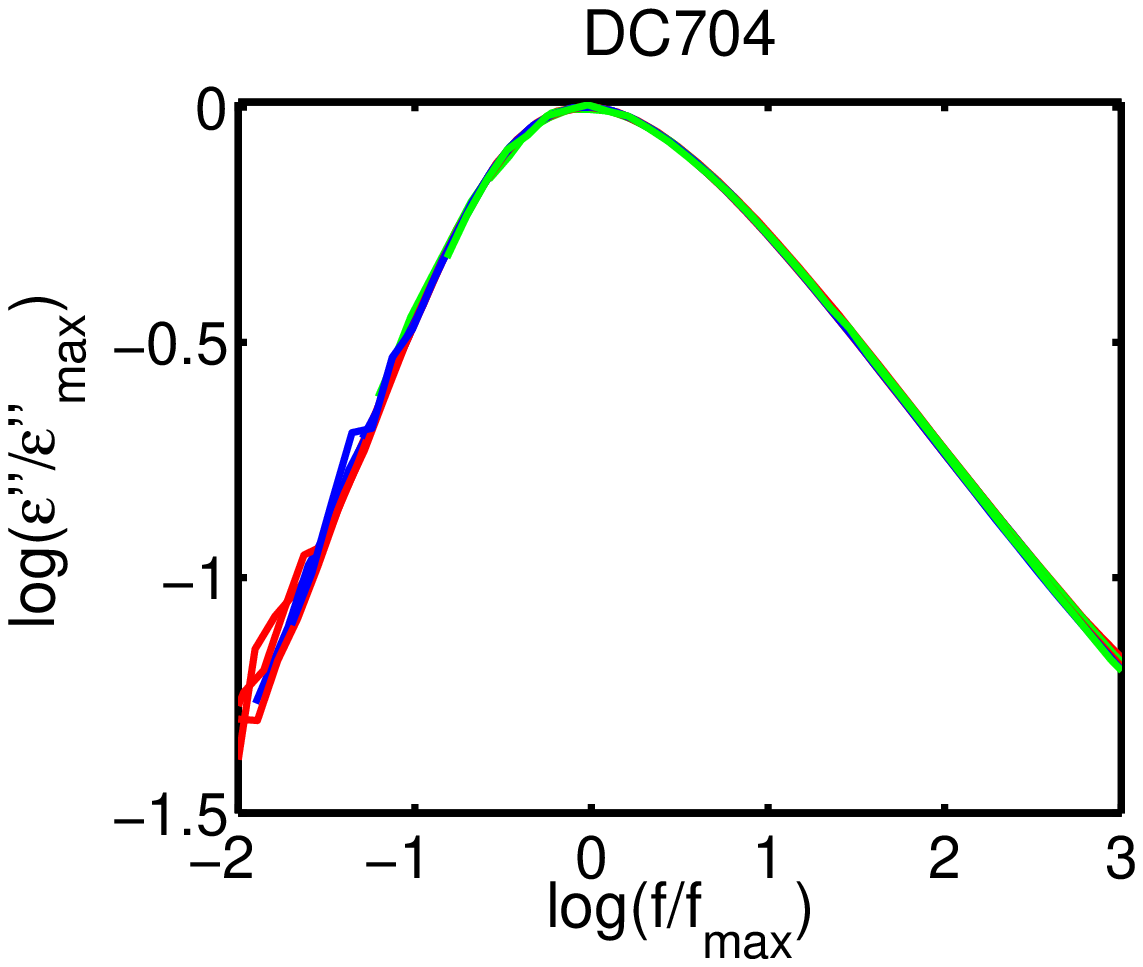}
\end{minipage}
\begin{minipage}{0.3\textwidth}
\centering
\includegraphics[width=1\textwidth]{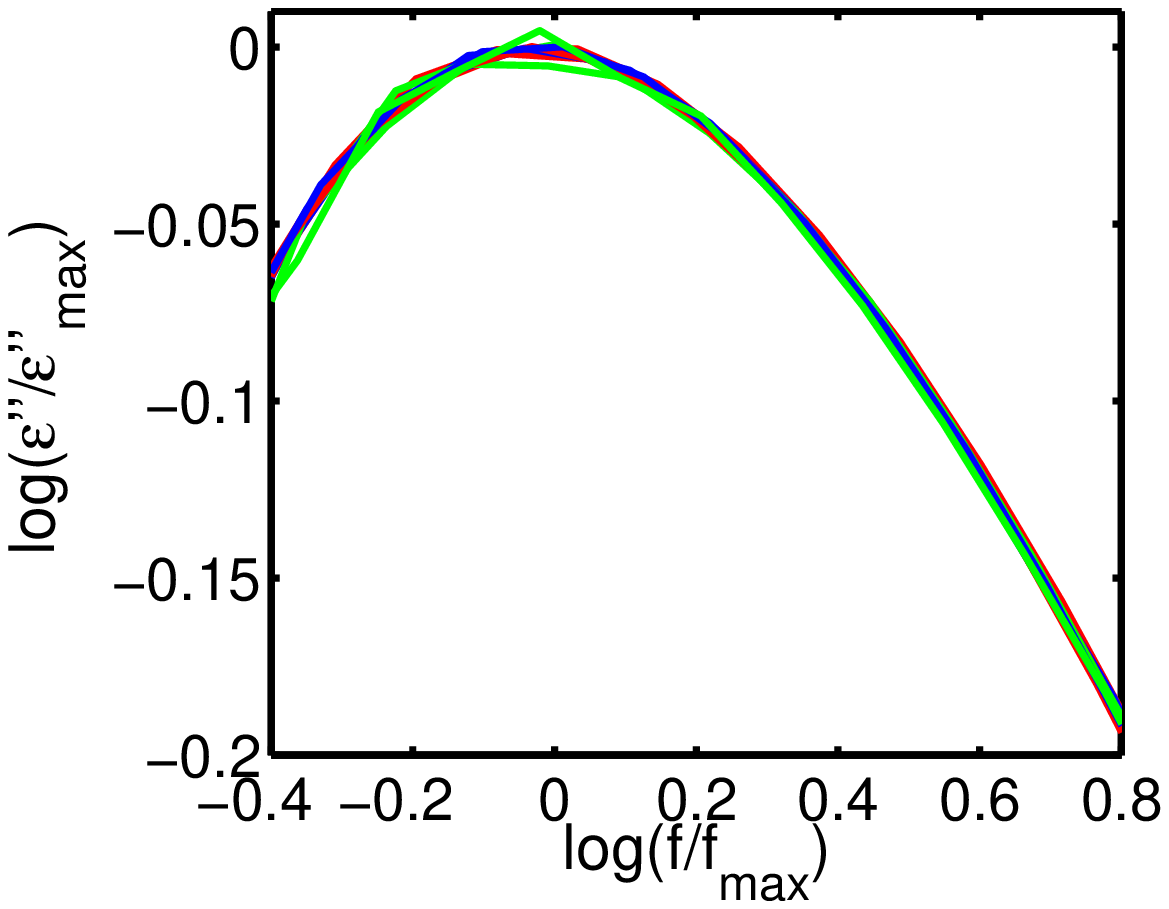}
\end{minipage}
\caption{The normalized relaxation spectra for DC704. The three different colors show the three isochrones. The figure to the right is a zoom of the peak. The red spectra are measurements with log$(f_{m})\simeq$ 2, the blue spectra are measurements with log$(f_{m})\simeq$ 1, and the green spectra are measurements with log$(f_{m})\simeq$ 0.3.}
\label{DC704TTPS}
\end{figure}

\begin{figure}
\begin{minipage}{0.3\textwidth}
\centering
\includegraphics[width=1\textwidth]{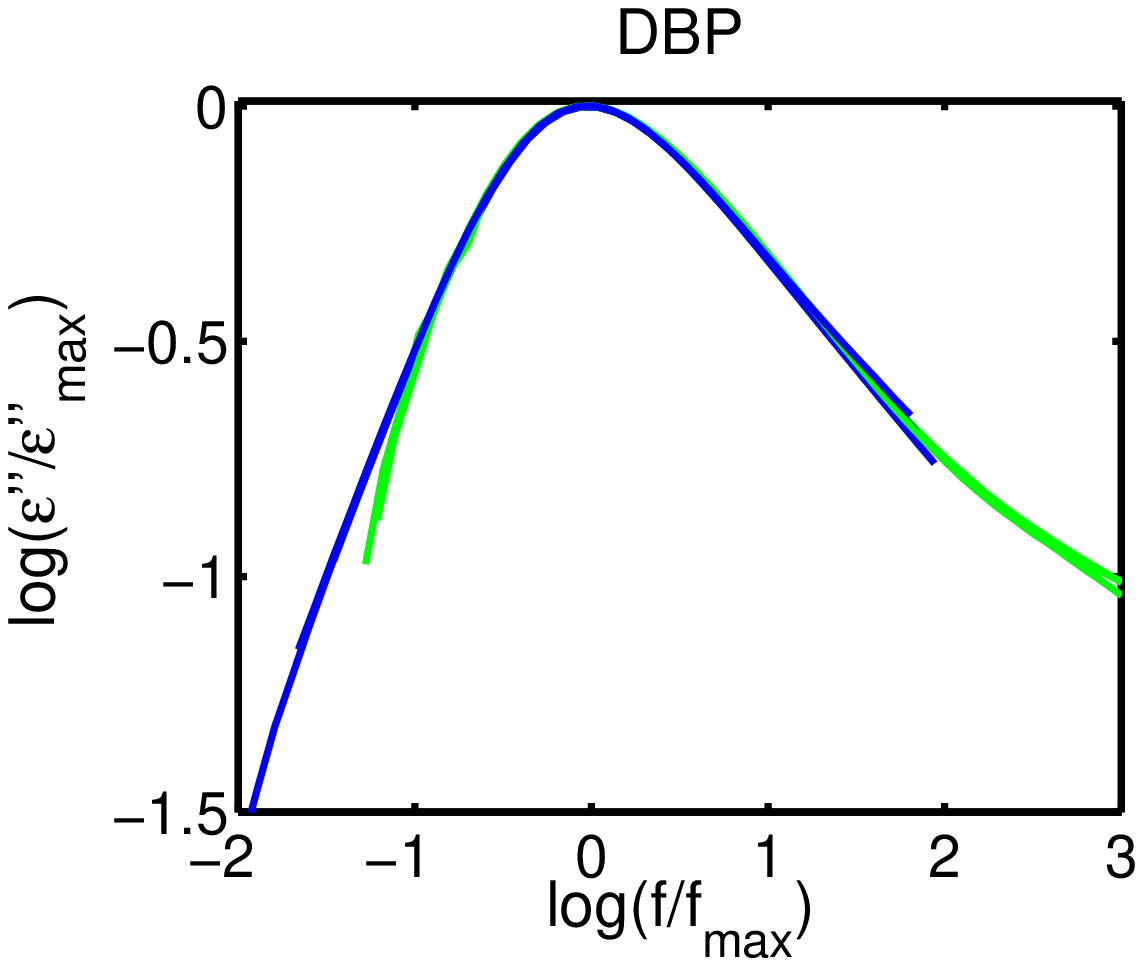}
\end{minipage}
\begin{minipage}{0.3\textwidth}
\centering
\includegraphics[width=1\textwidth]{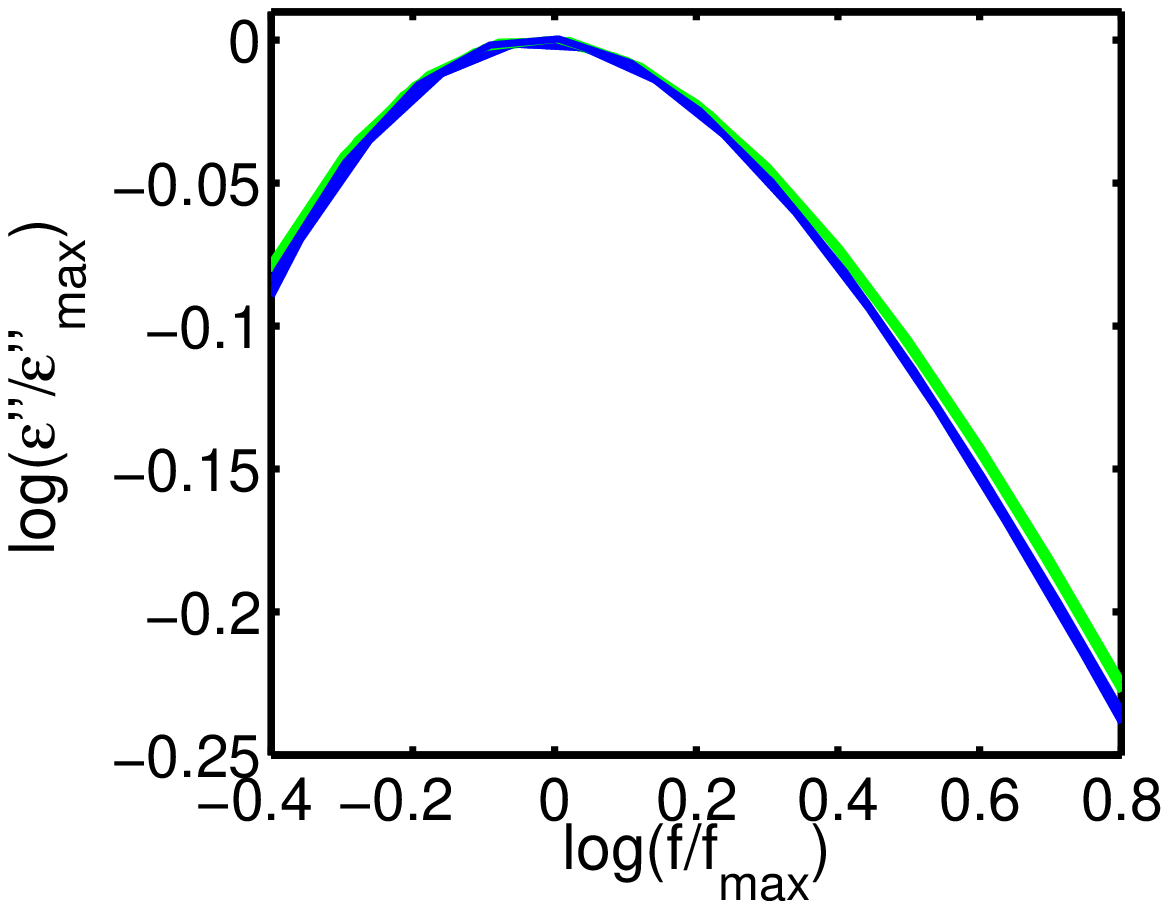}
\end{minipage}
\caption{The normalized relaxation spectra for DBP. The two different colors show the two isochrones. The figure to the right is a zoom of the peak. The blue spectra are measurements with log$(f_{m})\simeq$ 4.1, and the green spectra are measurements with log$(f_{m})\simeq$ 2.6.}
\label{DBPTTPS}
\end{figure}

For each liquid, all the used raw data are normalized, and plotted in the same figure (the Figs. \ref{126HTTTPS}, \ref{glycerolTTPS}, \ref{5PPETTPS}, \ref{DEPTTPS}, \ref{DC704TTPS}, and \ref{DBPTTPS}). Measurements that are from same isochrone have the same color. The figures thereby provide a qualitative picture of how close the relaxation spectra on the same isochrone are to each other, as well as giving a qualitative idea of whether the relaxation spectra are more equal on the isochrones than in general.

From the figures, it is clear that 1,2,6-HT obeys isochronal superposition better than TTPS. This is also the case for glycerol and DBP although more clearly with 1,2,6-HT. With 5PPE, DEP, and DC704 it also seems like this is the case, but it is difficult to quantify.

\subsection{Isochrones from a theoretical point of view}

From a theoretical point of view, by reference to Newton's laws of motion, an isochrone should be defined by requiring constant {\it reduced} relaxation time \cite{gnan2009}. The reduced relaxation time, however, is constructed by multiplying the relaxation time by the square root of temperature and the cubic root of density; the variation of this factor is entirely insignificant compared to the relaxation time variation for glass-forming liquids with relaxation times much longer than picoseconds. For this reason we stick to the standard definition of an isochrone.

\subsection{Density data}

Density measurements are available for DBP and glycerol in the literature. We use the
data from Ref. \onlinecite{bridgman1932}. The data is fitted to the 
Tait-equation in this form:

\begin{equation}
\label{taiteq}
V_{sp}=v_0\cdot e^{\alpha_0\cdot T}\cdot \left( 1-C\cdot\text{ln} \left(1+\frac{P}{b_0\cdot e^{-b_1\cdot T}} \right) \right)
\end{equation}
where $T$ is the temperature in Celsius and $P$ is the
pressure is in MPa. The parameters from the fits are given i table \ref{tait}.\\

PVT-measurements for DC704 and for 5PPE are made in
relation to Refs. \onlinecite{gundermann2011} and \onlinecite{gundermann2012}, where the parameters to the Tait-equation are found.
In relation to this investigation, we have performed PVT-measurements on 1,2,6-HT and DEP.
These data are also fitted to the Tait-equation. The parameters from these fit are also given in table \ref{tait}.

\begin{table}
\begin{tabular}{|l|r|r|r|r|r|l|}
\hline
\tiny{Liquid}   & \tiny{$v_0$} & \tiny{$a_0$} & \tiny{$C$} & \tiny{$b_0$} & \tiny{$b_1$} & \tiny{Ref.} \\
\hline
\tiny{1,2,6-HT} & \tiny{$0.894$} & \tiny{$4.9\cdot10^{-4}$} & \tiny{$0.085$} & \tiny{$269$} & \tiny{$0.0029$} & \tiny{This work} \\
\hline
\tiny{Glycerol} & \tiny{$0.808$} & \tiny{$4.8\cdot10^{-4}$} & \tiny{$0.112$} & \tiny{$519$} & \tiny{$0.0018$} & \tiny{Ref. \onlinecite{bridgman1932}} \\
\hline
\tiny{5PPE  }   & \tiny{$0.822$} & \tiny{$6.5\cdot10^{-4}$} & \tiny{$0.095$} & \tiny{$285$} & \tiny{$0.0043$} & \tiny{Ref. \onlinecite{gundermann2012}} \\
\hline
\tiny{DEP   }   & \tiny{$0.874$} & \tiny{$7.4\cdot10^{-4}$} & \tiny{$0.103$} & \tiny{$235$} & \tiny{$0.0046$} & \tiny{This work} \\
\hline
\tiny{DC704 }   & \tiny{$0.920$} & \tiny{$7.1\cdot10^{-4}$} & \tiny{$0.088$} & \tiny{$188$} & \tiny{$0.0048$} & \tiny{Ref. \onlinecite{gundermann2012}} \\
\hline
\tiny{DBP }    & \tiny{$0.938$} & \tiny{$7.9\cdot10^{-4}$} & \tiny{$0.093$} & \tiny{$189$} & \tiny{$0.0050$} & \tiny{Ref. \onlinecite{bridgman1932}} \\
\hline
\end{tabular}
\caption{The parameters for the Tait-equation (Eq. \ref{taiteq}) for the different liquids.}
\label{tait}
\end{table}

\subsection{Data of the selected measurements}\label{datas}
\begin{tabular}{cccccc}
\begin{minipage}[t]{2.5cm}
\textbf{1,2,6-HT}\\
\begin{tabular}{|r|r|r|}
\hline
\tiny{log$(f_{m})$} & \tiny{T (K)}  & \tiny{p (MPa)}\\
\hline
\tiny{4.05} & \tiny{240.5} & \tiny{99} \\
\tiny{4.03} & \tiny{243.0} & \tiny{150} \\
\tiny{4.02} & \tiny{245.3} & \tiny{199} \\
\tiny{4.06} & \tiny{248.0} & \tiny{248} \\
\tiny{4.07} & \tiny{250.5} & \tiny{298} \\
\hline
\tiny{3.53} & \tiny{235.6} & \tiny{101} \\
\tiny{3.54} & \tiny{238.0} & \tiny{150} \\
\tiny{3.54} & \tiny{240.5} & \tiny{198} \\
\tiny{3.52} & \tiny{242.8} & \tiny{250} \\
\tiny{3.56} & \tiny{245.3} & \tiny{298} \\
\tiny{4.65} & \tiny{248.5} & \tiny{347} \\
\hline
\tiny{2.96} & \tiny{235.6} & \tiny{198} \\
\tiny{2.95} & \tiny{237.9} & \tiny{251} \\
\tiny{2.99} & \tiny{240.5} & \tiny{301} \\
\tiny{3.04} & \tiny{243.2} & \tiny{351} \\
\tiny{3.05} & \tiny{245.3} & \tiny{399} \\
\hline
\end{tabular}
\end{minipage}
&
\begin{minipage}[t]{2.5cm}
\textbf{Glycerol}\\
\begin{tabular}{|r|r|r|}
\hline
\tiny{log$(f_{m})$} & \tiny{T (K)}  & \tiny{p (MPa)}\\
\hline
\tiny{4.94} & \tiny{242.4} & \tiny{100} \\
\tiny{4.95} & \tiny{244.7} & \tiny{151} \\
\tiny{4.94} & \tiny{246.3} & \tiny{200} \\
\tiny{4.96} & \tiny{248.6} & \tiny{249} \\
\tiny{4.96} & \tiny{250.4} & \tiny{298} \\
\hline
\tiny{4.52} & \tiny{237.5} & \tiny{100} \\
\tiny{4.54} & \tiny{240.0} & \tiny{152} \\
\tiny{5.52} & \tiny{241.4} & \tiny{199} \\
\tiny{4.52} & \tiny{243.5} & \tiny{249} \\
\tiny{4.53} & \tiny{245.3} & \tiny{300} \\
\hline
\tiny{4.05} & \tiny{236.6} & \tiny{198} \\
\tiny{4.07} & \tiny{238.6} & \tiny{249} \\
\tiny{4.06} & \tiny{240.4} & \tiny{299} \\
\hline
\end{tabular}
\end{minipage}
&
\begin{minipage}[t]{2.5cm}
\textbf{5PPE}\\
\begin{tabular}{|r|r|r|}
\hline
\tiny{log$(f_{m})$} & \tiny{T (K)}  & \tiny{p (MPa)}\\
\hline
\tiny{3.98} & \tiny{271.9} & \tiny{0.1} \\
\tiny{3.96} & \tiny{295.3} & \tiny{97} \\
\tiny{3.96} & \tiny{317.9} & \tiny{199} \\
\hline
\tiny{3.05} & \tiny{267.1} & \tiny{0.1} \\
\tiny{3.05} & \tiny{290.4} & \tiny{99} \\
\tiny{3.06} & \tiny{312.0} & \tiny{198} \\
\tiny{3.09} & \tiny{332.4} & \tiny{300} \\
\hline
\tiny{0.30} & \tiny{255.6} & \tiny{0.1} \\
\tiny{0.21} & \tiny{277.7} & \tiny{101} \\
\tiny{0.17} & \tiny{297.3} & \tiny{199} \\
\tiny{0.26} & \tiny{316.9} & \tiny{301} \\
\hline
\end{tabular}
\end{minipage}
&
\begin{minipage}[t]{2.5cm}
\textbf{DEP}\\
\begin{tabular}{|r|r|r|}
\hline
\tiny{log$(f_{m})$} & \tiny{T (K)}  & \tiny{p (MPa)}\\
\hline
\tiny{4.88} & \tiny{237.3} & \tiny{149} \\
\tiny{4.93} & \tiny{245.4} & \tiny{200} \\
\tiny{4.89} & \tiny{251.3} & \tiny{248} \\
\tiny{4.88} & \tiny{258.2} & \tiny{298} \\
\tiny{4.85} & \tiny{263.9} & \tiny{347} \\
\tiny{4.91} & \tiny{270.9} & \tiny{398} \\
\hline
\tiny{4.34} & \tiny{240.5} & \tiny{200} \\
\tiny{4.26} & \tiny{246.3} & \tiny{251} \\
\tiny{4.31} & \tiny{253.2} & \tiny{299} \\
\tiny{4.30} & \tiny{259.1} & \tiny{350} \\
\tiny{4.25} & \tiny{265.0} & \tiny{401} \\
\hline
\tiny{3.37} & \tiny{239.5} & \tiny{248} \\
\tiny{3.40} & \tiny{246.3} & \tiny{299} \\
\tiny{3.39} & \tiny{252.2} & \tiny{351} \\
\tiny{3.40} & \tiny{258.2} & \tiny{400} \\
\hline
\tiny{2.43} & \tiny{240.4} & \tiny{300} \\
\tiny{2.36} & \tiny{245.2} & \tiny{348} \\
\tiny{2.37} & \tiny{251.4} & \tiny{400} \\
\hline
\tiny{1.69} & \tiny{236.4} & \tiny{300} \\
\tiny{1.66} & \tiny{241.3} & \tiny{347} \\
\tiny{1.70} & \tiny{247.5} & \tiny{399} \\
\hline
\end{tabular}
\end{minipage}
&
\begin{minipage}[t]{2.5cm}
\textbf{DC704}\\
\begin{tabular}{|r|r|r|}
\hline
\tiny{log$(f_{m})$} & \tiny{T (K)}  & \tiny{p (MPa)}\\
\hline
\tiny{1.89} & \tiny{253} & \tiny{104.3} \\
\tiny{2.22} & \tiny{263} & \tiny{155.3} \\
\tiny{1.84} & \tiny{283} & \tiny{255.9} \\
\hline
\tiny{0.90} & \tiny{253} & \tiny{132.0} \\
\tiny{1.05} & \tiny{263} & \tiny{178.7} \\
\tiny{1.11} & \tiny{283} & \tiny{274.4} \\
\hline
\tiny{0.21} & \tiny{253} & \tiny{144.5} \\
\tiny{0.33} & \tiny{263} & \tiny{192.4} \\
\tiny{0.23} & \tiny{283} & \tiny{294.7} \\
\hline
\end{tabular}
\end{minipage}
&
\begin{minipage}[t]{2.5cm}
\textbf{DBP}\\
\begin{tabular}{|r|r|r|}
\hline
\tiny{log$(f_{m})$} & \tiny{T (K)}  & \tiny{p (MPa)}\\
\hline
\tiny{4.06} & \tiny{206} & \tiny{0} \\
\tiny{4.16} & \tiny{219.3} & \tiny{108} \\
\tiny{4.19} & \tiny{236.3} & \tiny{251} \\
\hline
\tiny{2.61} & \tiny{206} & \tiny{85} \\
\tiny{2.68} & \tiny{219.3} & \tiny{200} \\
\tiny{2.40} & \tiny{236.3} & \tiny{389} \\
\hline
\end{tabular}
\end{minipage} \\
\end{tabular}

\pagebreak

\end{document}